\documentstyle{article}
\begin{document}\large\rm
\baselineskip=0.75cm

\title{\bf Chaos and Asymptotical Stability in  Discrete-time Neural Networks}

\author{Luonan Chen \\{The Kaihatsu Computing Service Center, 2-2-18, Fukagawa, Koto-Ku, Tokyo 135, Japan} \and Kazuyuki Aihara \\{The University of Tokyo,
 7-3-1 Hongo, Bunkyo-Ku, Tokyo 113, Japan}}
\date{}
\maketitle

\clearpage
\setcounter{page}{1}
\pagestyle{headings}
This paper will be published in {\bf Physica D.}\\[1.2cm]

Title:

\begin{center}\LARGE
{\bf Chaos and Asymptotical Stability in   Discrete-time  Neural Networks}
\end{center}

\vspace*{1.0cm}

Authors:

\begin{center}
Luonan Chen\\
Power System Department,\\
The Kaihatsu Computing Service Center Ltd.,\\
2-2-18, Fukagawa,Koto-Ku, Tokyo 135, Japan\\
TEL: +81-3-3642-9771 \  FAX: +81-3-3642-9796 \ Email: chen@kcc.co.jp
\end{center}

\vspace*{0.3cm}

\begin{center}
Kazuyuki Aihara\\
Department of Mathematical Engineering and Information Physics,\\
Faculty of Engineering,\\
The University of Tokyo,\\
7-3-1 Hongo, Bunkyo-Ku, Tokyo 113, Japan\\
TEL: +81-3-3812-2111 \ FAX: +81-3-5689-5752 \ Email: Aihara@sat.t.u-tokyo.ac.jp
\end{center}

\vspace*{1.2cm}
Acknowledgments: The authors wish to express their thanks to S.Amari for his encouragement. This research was partially supported by a Grant-in-Aid for Scientific Research on Priority Areas from the Ministry of Education, Science and Culture of Japan.

\vspace*{0.3cm}
{\it Running title: Chaos and Asymptotical Stability in Neural Networks}

\clearpage  
\pagestyle{myheadings}
\markright{Abstract}

\begin{center}
{\bf Abstract}
\end{center}

This paper aims to theoretically prove by applying Marotto's Theorem that both 
transiently chaotic neural networks (TCNN) and discrete-time recurrent neural
 networks (DRNN)  have  chaotic structure. A significant property of TCNN and 
DRNN  is that they have only one fixed point, when absolute values of the 
self-feedback connection weights in TCNN and the difference time in DRNN are 
sufficiently large. We show that this unique fixed point can actually evolve 
into a snap-back repeller which generates chaotic structure, if several
 conditions are satisfied. On the other hand, by using the Lyapunov functions,
 we also derive sufficient conditions on asymptotical stability for symmetrical versions of both TCNN and DRNN, under which TCNN and DRNN asymptotically
 converge to a fixed point. Furthermore, generic bifurcations are also
 considered in this paper.  Since both of TCNN and DRNN are not special but
 simple and general,  the obtained theoretical results hold for a wide class of discrete-time neural networks. To demonstrate the theoretical results of this
 paper better, several numerical simulations are provided as illustrating 
examples.

\vspace*{5mm}

{\bf Keywords-} Neural network, Chaos, Snap back repeller, Simulated annealing, Asymptotical stability. 

\clearpage
\pagestyle{myheadings}
\markright{List of Symbols}

\begin{center}
List of Symbols
\end{center}

\vspace*{2cm}

\begin{tabular}{ll}
$x_{i} $  &: output of neuron $i$ \\
$y_{i} $  &: internal state of neuron $i$ \\
$\omega_{ij}$  &: connection weight from neuron $j$ to neuron $i$\\
$I_{i}$   &: input bias of neuron $i$ \\
$k$       &: damping factor of nerve membrane \\
$\omega_{ii}, \omega_{ii}(t) $  &: self-feedback connection weight\\
$W$ & : connection weight matrix\\
$\epsilon$ &: steepness parameter of output function($\epsilon >0$)\\
$f$        &: Lyapunov function\\
$\Delta t$ &: time step of the Euler method \\
$\beta $   &: damping factor for $\omega_{ii}$\\
$\lambda_{p} $ &: eigenvalues of matrix\\
$\lambda_{min} $ &: the smallest eigenvalue of matrix $W$\\
$\lambda_{max} $ &: the largest eigenvalue of matrix $W$\\
$I_{0}$  &: self-recurrent bias\\
$F^{t}$ & : the composition of $F$ with itself $t$ times\\
$F_{Y}(Y)$ & :the Jacobian matrix of $F$ at the point $Y \in  \mbox{ \boldmath$  R $} ^{n}$\\
$det(F_{Y}(Y))$ & : determinant of $F_{Y}(Y)$\\
$B(Y;r)$ & : closed ball in $ \mbox{ \boldmath$  R $}^{n}$ of radius $r$
 centered at the point $Y$\\
$B^{0}(Y;r)$ &: interior of $B(Y;r)$\\
$Y^{T}$ &: transpose of the column vector $Y$\\
$\| \cdot \|$ & : the usual Euclidean norm of  a vector or the induced matrix
 norm of a matrix\\
$E_{0}$ &: $=(0,...,0)^{T}$\\
$ E_{1}$ &: $ = (1,...,1)^{T}$\\
{$ \mbox{ \boldmath$1$} $} &: the identity matrix\\
$S$ &:  region defined as $S= \{ (x_{1},...,x_{n}) | 0 \leq x_{i} \leq 1; 
i=1,...,n \}$\\
$S_{V}$ &: vertices defined as $S_{V}= \{(x_{1},...,x_{n}) | x_{i}=0 \  
\mbox{or} \ 1; i=1,...,n \}$\\
$S_{I}$ &: internal points defined as $S_{I} = \{ (x_{1},...,x_{n}) 
| 0 < x_{i} <1; i=1,...,n  \}$\\
$S_{B}$ &: boundary points defined as $S_{B}= S -S_{I} -S_{V}$\\
$x \propto y$ &:  $x$ is directly proportional to $y$\\
{$ \mbox{ \boldmath$0$} $} &: the $n \times n$ zero matrix\\
 \end{tabular}

\clearpage
\pagestyle{headings}

\section{Introduction}
 
Up to now, artificial neural networks  (e.g. Amari, 1972, Hopfield, 1982;
 Hopfield \& Tank, 1985; Rumelhart, et al., 1986; Wells, 1992; Peterson \& Anderson, 1989; Peterson \& Soderberg, 1993)
 have been widely applied to various information processing problems with considerable success. In particular, the transiently chaotic neural networks (TCNN) 
 with chaotic simulated annealing (CSA) have  been shown to be a  promising tool for combinatorial optimization \cite{Chen93,Chen94,Chen95a,Chen95b}. Since TCNN has rich dynamics with various coexisting attractors, not only of fixed points but also periodic and even chaotic attractors, it can be expected to have high ability of escaping from local minima and searching for globally optimal or near-optimal solutions. However, although it has numerically been observed that the numerical chaos with positive Lyapunov exponents exists in TCNN when the temperature, or the self-feedback connection weight is sufficiently large,
 it has not yet theoretically been proved that TCNN does have chaotic structure.

On the other hand,  there are a variety of mathematical definitions of chaos, such as geometrically constructed chaos (or geometric chaos) including Horseshoe, Solenoid, and Plykin systems, formal chaos (or topological chaos) with positive topological entropy where shift mapping is included as a subsystem, and numerical chaos in which the maximal  Lyapunov exponent is numerically observed to be positive, and so on \cite{Komuro}.  The definition of chaos widely used in discrete-time dynamical systems is the Li-York chaos which was proposed by Li and York and which actually corresponds to topological chaos \cite{Li}. In 1975, Li and York derived a surprisingly simple condition ensuring the existence of chaos in one-dimension systems, that is, "Period three implies chaos" \cite{Li}. In 1978, Marotto extended the Li-York's work to multi-dimensional systems by defining snap-back repellers, that is, "Snap-back repellers imply chaos in $ \mbox{ \boldmath$  R $} ^{n}$" 
\cite{Marotto}.  In 1979, Shiraiwa and Kurata further generalized the theorem of Marotto to include not only repellers but also saddle points \cite{Shiraiwa}. Shiraiwa and Kurata also showed that their theorem was  a generalization of Smale's results \cite{Smale} on transversal homoclinic points in some sense. 

On  chaos in  discrete-time systems obtained by temporally discretizing continuous-time systems, in 1979 Yamaguti and Matano proved that there exists  chaotic structure in some class of one-dimensional difference equations with two fixed points when the difference time is sufficiently large, by using Li-York's Theorem \cite{Yamaguti}. In 1982, Hata extended the works of Yamaguti and Matano to multi-dimensional difference equations with two fixed points, by using Marotto's Theorem \cite{Hata}. In 1983, Ushio and Hirai further showed that there exists chaotic behavior in nonlinear sampled-data control systems with two fixed points when the sampling period is sufficiently large \cite{Ushio83,Ushio85}, by using Marotto's Theorem and Shiraiwa-Kurata's Theorem.

Motivated by the previous works, this paper aims to theoretically prove that both TCNN and more general discrete-time recurrent neural networks (DRNN) have chaotic structure by applying Marotto's Theorem. A significant property of TCNN and DRNN  is that they have only one fixed point, when absolute values of the self-feedback connection weights in TCNN and the difference time in DRNN are sufficiently large. We show that this unique fixed point can actually evolve into  a snap-back repeller which generates chaotic structure, if several conditions are satisfied. 

For the aspects of asymptotical stability, on the other hand, a number of progresses have been made for continuous-time neural networks. For instance, it has been proven that continuous-time neural networks with symmetrical connections  always converge to a fixed point\cite{Hopfield84,Hopfield85}. However, theoretical results of continuous-time neural networks do not necessarily hold for discrete-time neural networks with continuous-states because analysis of discrete-time neural networks, which also depends on the updating schemes such as synchronous updating and asynchronous updating, is quite different from that of continuous-time neural networks. 
 For some special forms of discrete-time neural networks with continuous-states which are widely used in the analysis of mean field equations \cite{Peterson93},
 Marcus and Westervelt indicated in 1989 that there exist only stable fixed points and stable period-two cycles \cite{Marcus}.
Although it has numerically been observed that TCNN with CSA converges to a fixed point attractor if the absolute value of self-connections is sufficient small, it remains theoretically unclear what conditions are sufficient for the convergence and which fixed points are asymptotically stable in TCNN.
 
In this paper, we also aim to analyse  discrete-time neural networks of both TCNN and DRNN, and to investigate the asymptotical stability as well as the relations between fixed points and local minima of the Lyapunov functions. In other words, 
by using Lyapunov functions, we derive sufficient conditions on asymptotical stability for several symmetrical versions of both TCNN and DRNN, under which these networks asymptotically converge to a fixed point.  
We show that the convergent properties of discrete-time recurrent neural networks are quite different from those of continuous-time recurrent neural networks although many of them are similar in terms of asymptotical stability and locations of fixed points. Furthermore, the relationships between the asymptotically stable points and the minima of the Lyapunov functions are also investigated for both asynchronously and synchronously updating schemes, by which TCNN and DRNN can be applied to combinatorial optimization problems. In addition, we  also give conditions of generic bifurcations, i.e., the fold and flip bifurcations, and explain the bifurcation structure of chaotic simulated annealing. Since both of TCNN and DRNN are not special but simple and general,  the obtained theoretical results hold for a wide class of discrete-time neural networks. To demonstrate the theoretical results of this paper better, several numerical simulations are provided as illustrating examples.!

This paper is organized as follows:  In the next section, Marotto's Theorem and several definitions are briefly described. In section three, we first give a general form of TCNN and then prove convergence theorems for several symmetrical versions of TCNN.  In addition, bifurcation conditions for flip and fold bifurcations are also given in section three. We show that all of the theoretical results in section three also hold for chaotic neural networks (CNN) \cite{Aihara}. Section four describes one  theorem for the existence of a fixed point in TCNN with the sufficiently large absolute values of self-feedback connection weights, and  gives sufficient conditions for chaos in TCNN.  Asymptotical stability conditions and the proof of existence of chaos in DRNN are given in  section five and six, respectively.  Finally, we give some general remarks to conclude this paper.

\clearpage
\pagestyle{myheadings}
\markright{2 NOTATION AND MAROTTO'S THEOREM}

\section{Notation and Marotto's Theorem}

We use the following notation, similar to that of Marotto \cite{Marotto}
 throughout the paper. Let us define a discrete-time system as, 

\begin{equation}
Y(t+1) = F(Y(t)) \label{e1}
\end{equation}

where, $Y(t) \in \mbox{ \boldmath$  R $}^{n}$, $(t=1,2,...)$  and $F \in C^{1}:  \mbox{ \boldmath$  R  $}^{n} \rightarrow  \mbox{ \boldmath$  R  $}^{n}$.

Let $F^{t}$ denote the composition of $F$ with itself $t$ times, then a point $Y$ is a $p$-periodic point if $F^{p}(Y) = Y$ but $F^{t}(Y) \neq Y$ for  $1 \leq t < p$. If $p = 1$, that is, $F(Y) = Y$, $Y$ is called a fixed point.
Let $F_{Y}(Y)$ and $det(F_{Y}(Y))$
be the Jacobian matrix of $F$ at the point $Y \in  \mbox{ \boldmath$  R  $}^{n}$ and its determinant, respectively. Let $B(Y;r)$ be  a closed ball in $ \mbox{ \boldmath$  R  $} ^{n}$ of radius $r$ centered at the point $Y$, and $B^{0}(Y;r)$ its interior. Furthermore, let $Y^{T}$ be the transpose of the column vector $Y$.  Define $\| \cdot \|$ to be the usual Euclidean norm of  a vector or the induced matrix norm of a matrix.
 Define $E_{0}=(0,...,0)^{T}, E_{1} = (1,...,1)^{T}$, and {$ \mbox{ \boldmath$1$} $} to be the identity matrix . Define the region $S= \{ (x_{1},...,x_{n}) | 0 \leq x_{i} \leq 1; i=1,...,n \}$. Then $S$ can be divided into vertices $S_{V} = \{(x_{1},...,x_{n}) | x_{i}=0 \  \mbox{or} \ 1; i=1,...,n \}$, internal points $S_{I} = \{ (x_{1},...,x_{n}) | 0 < x_{i} <1; i=1,...,n  \}$ and boundary points $S_{B}= S -S_{I} -S_{V}$, respectively. $x \propto y$ means that $x$ is directly proportional to $y$.
{$ \mbox{ \boldmath$0$} $} is the $n \times n$ zero matrix. A small letter stands for a scalar, and a capital letter represents a vector, a matrix or a set in this paper.

\newtheorem{definition}{Definition}[section]
\begin{definition}[Marotto, 1978]
A fixed point $Y^{*}$ is said to be a snap-back repeller of eqn.(\ref{e1}), if there exists a real number $r \   (>0)$ and $Y^{0} \in B(Y^{*}; r)$ with $Y^{0} \neq Y^{*}$ such that all eigenvalues of $F_{Y}(Y)$ exceed the unity in norm for all $Y \in B(Y^{*}; r)$ and $F^{m}(Y^{0}) = Y^{*}$ with $det( F_{Y}^{m}(Y^{0}) ) \neq 0$ for some positive integer $m$.
\end{definition}

\newtheorem{theorem}{Theorem}[section]
\begin{theorem}[Marotto, 1978]
\label{t21}
If eqn.(\ref{e1}) has a snap-back repeller then the system of eqn.(\ref{e1}) is chaotic.  In other words, there exist,

\begin{enumerate}
\item a positive integer $n$ such that for each integer $p \geq n$, eqn.(\ref{e1}) has $p-$periodic points;
\item a scrambled set, i.e, an uncountable set $L$ containing no periodic points such that 
\begin{enumerate}
     \item $F(L) \subset L$
     \item for every $Y \in L$ and any periodic point $X$ of eqn.(\ref{e1}) 

          \[ { \displaystyle \limsup_{t \rightarrow \infty }} \| F^{t}(Y)-F^{t}(X) \| > 0 \]
     \item for every $X,Y \in L$ with $X \neq Y$ ;

           \[ { \displaystyle \limsup_{t \rightarrow \infty }} \| F^{t}(Y)-F^{t}(X) \| > 0 \]
     \end{enumerate}
\item an uncountable subset $L_{0}$ of $L$ such that for every $X,Y \in L_{0}$

          \[ { \displaystyle \liminf_{t \rightarrow \infty }} \| F^{t}(Y)-F^{t}(X) \| = 0 \]
\end{enumerate}

\end{theorem}

\clearpage
\pagestyle{myheadings}
\markright{3 ASYMPTOTICAL STABILITY OF TCNN}

\section{
Asymptotical Stability and Local Bifurcations of Transiently Chaotic Neural Networks}

In this section, first the transiently chaotic neural network (TCNN) is briefly summarized. Then, several convergence theorems for symmetrical versions of TCNN are derived. Besides, the relations between asymptotically stable points of TCNN and local minimal points of the Lyapunov function are also investigated.

\subsection{A model of the transiently chaotic neural network}

The model of the transiently chaotic neural network (TCNN) can be described in terms of scalar variables as follows \cite{Chen93,Chen94,Chen95a,Chen95b}:

\begin{eqnarray}
x_{i}(t) = \frac{1}{1+e^{-  y_{i}(t) / \epsilon }}  \label{e2}\\
y_{i}(t+1) = k y_{i}(t) + \sum_{j=1}^{n} \omega_{ij}x_{j}(t) + a_{i}- \omega_{ii}(t)a_{0i} 
\label{e3} \\
| \omega_{ii}(t+1)|=(1-\beta)| \omega_{ii}(t)|  \label{e4}\\ 
 (i=1,...,n) \nonumber
 \end{eqnarray}
 
where, $\omega_{ii}= \omega_{ii}(t)$ is a variable depending on eqn.(\ref{e4}) and all variables and parameters are real numbers, and

\begin{tabular}{ll}
$x_{i} $  &: output of neuron $i$\\
$y_{i} $  &: internal state of neuron $i$\\
$\omega_{ij}$  &: connection weight from neuron $j$ to neuron $i$\\
$a_{i}$   &: input bias of neuron $i$\\
$k$       &: damping factor of nerve membrane\\
$\epsilon$ &: steepness parameter of the output function ($\epsilon >0$)\\
$\omega_{ii}$  &: self-feedback connection weight, where $\omega_{ii}= \omega_{ii}(t)$ \\
$ \beta$ &: damping factor of the time-dependent $\omega_{ii}(t)$ ($0 \le \beta \le 1$)\\
$a_{0i}$ &: self-recurrent bias of neuron $i$.
\end{tabular}

Since we investigate sufficient conditions of chaos and asymptotical stability for TCNN with temporally constant $\omega_{ii}$, $\omega_{ii}$ is fixed by dropping eqn.(\ref{e4}) in the following analysis.Therefore, TCNN can be rewritten in terms of vector variables as follows.

 \begin{equation}
Y(t+1) = F(Y(t)) = k Y(t) + WX(Y(t)) + I - diag(\omega_{11},...,\omega_{nn})
 I_{0}       
\label{e5}
 \end{equation}
 
where, $diag(\omega_{11},...,\omega_{nn})$ is a diagonal matrix and 

\[
Y(t) = \left[
\begin{array}{c}
y_{1}(t) \\ \vdots \\ y_{n}(t)
\end{array}
\right]
, \ \ \ 
X(Y(t)) = \left[
\begin{array}{c}
x_{1}(t) \\ \vdots \\ x_{n}(t)
\end{array}
\right] 
=
 \left[
\begin{array}{c}
\frac{1}{1+e^{-  y_{1}(t) / \epsilon }} \\ \vdots \\ \frac{1}{1+e^{-  y_{n}(t) / \epsilon }}
\end{array}
\right]
\]

\[
I = \left[
\begin{array}{c}
a_{1} \\ \vdots \\ a_{n}
\end{array}
\right]
, \ \ \ 
I_{0} = \left[
\begin{array}{c}
a_{01} \\ \vdots \\ a_{0n}
\end{array}
\right] 
, \ \ \
W = \left[
\begin{array}{cccc}
\omega_{11} & \omega_{12} & \cdots & \omega_{1n} \\
\omega_{21} & \omega_{22} & \cdots & \omega_{2n} \\
\vdots  & \vdots      & \vdots & \vdots      \\
\omega_{n1} & \omega_{n2} & \cdots & \omega_{nn} 
\end{array}
\right]
\]

Obviously, $F$ in eqn.(\ref{e5}) is a $C^{1}$ class function for any bounded $Y$. Besides, $F$ can be viewed as an  iterated map or a difference equation or a neural network with discrete-time and continuous-state. If $I_{0}= E_{0}$, eqn.(\ref{e5}) coincides  with chaotic neural networks (CNN) \cite{Aihara,Chen95b}.

To analyse the asymptotical stability of $X(t)$, we rewrite eqn.(\ref{e5}) by explicitly expressing $X(t)$,

\begin{equation}
X(t+1) = \bar{F}(X(t)) = 
 \left[
\begin{array}{c}
\frac{1}{1+e^{-\frac{1}{\epsilon}[k \epsilon ln \frac{x_{1}(t)}{1-x_{1}(t)} 
 + \sum_{j=1}^{n} \omega_{1j}x_{j}(t) + a_{1}- \omega_{11} a_{01} ]}} \\
 \vdots \\ 
\frac{1}{1+e^{-\frac{1}{\epsilon}[k \epsilon ln \frac{x_{n}(t)}{1-x_{n}(t)} 
 + \sum_{j=1}^{n} \omega_{nj}x_{j}(t) + a_{n}- \omega_{nn} a_{0n} ]     }}
\end{array}
\right],
\label{e5a}
\end{equation}

where, $ 0 \leq x_{i}(t) \leq 1 $ for $i=1,...,n$.

\subsection{Asymptotical stability for TCNN}

Next, we first show several convergent theorems for  eqn.(\ref{e5a}), which ensure that the system of eqn.(\ref{e5a}) asymptotically converges to a stable fixed point, and then derive locally stable conditions by analysing the eigenvalues of the Jacobian matrix. Let us define two updating schemes, i.e., synchronously updating and asynchronously updating.  In synchronously updating, all neurons are updated in parallel, using only old values as input. In asynchronously updating, neurons are updated one by one, using fresh values of previously updated neurons \cite{Peterson93}.
Generally, asynchronously updating can further be divided into random-order asynchronously updating where neurons are chosen in random order, and fixed-order asynchronously updating where neurons are chosen in fixed-order, e.g., cyclic updating. In this paper, asynchronously updating means both random-order and fixed-order asynchronously updating. 

Let $\lambda_{min}, \lambda_{max}$ be the smallest eigenvalue and the largest eigenvalue  of the matrix $W$, respectively.  Define

\begin{equation}
f(X)=-\frac{1}{2}\sum^{n}_{ij} \omega_{ij}x_{i}x_{j} - \sum^{n}_{i}(-a_{0i}\omega_{ii} + a_{i})x_{i}
 - (k-1)\epsilon \sum^{n}_{i} \int_{0}^{x_{i}} ln \frac{x}{1-x} dx
\label{ely}
\end{equation}

 Note that $ \sum^{n}_{i} \int_{0}^{x_{i}} ln \frac{x}{1-x} dx =  \sum^{n}_{i} [ x_{i}lnx_{i} + (1-x_{i})ln(1-x_{i})]$.
Obviously, $f(X): \mbox{ \boldmath$  R $}^{n} \rightarrow \mbox{ \boldmath$  R $}$ is bounded for all $0 \leq x_{i} \leq 1 \ (i=1,...,n)$.

\begin{theorem}
\label{t31aa}
Assume $W=W^{T}$, and 
\begin{enumerate}
\item  $1/3 \geq k \geq 0$,  $4(1-k) \epsilon > - \lambda_{min}$\\
or
\item  $1 \geq k \geq 1/3$, $8k \epsilon > - \lambda_{min}$\\
or
\item  $ k > 1$, $8 \epsilon > - \lambda_{min}$.
\end{enumerate}
   Then  $X(t)$ of  the discrete-time system eqn.(\ref{e5a})  asymptotically converges to a fixed point, as far as eqn.(\ref{e5a}) is synchronously updated.
\end{theorem}

{\bf Proof of Theorem \ref{t31aa}}. We use the direct approach  based on a Lyapunov function, to prove the Theorem. 

{\sl For condition {\it 1}:}  Consider the change in $f$ between two discrete time steps under the assumption of synchronously updating, 

\begin{eqnarray}
f(X(t+1))-f(X(t)) 
=-\frac{1}{2} \sum^{n}_{ij} \omega_{ij} \Delta x_{i} \Delta x_{j} -k \epsilon \sum_{i}^{n} \Delta x_{i} [ ln \frac{x_{i}(t+1)}{1-x_{i}(t+1)} - ln \frac{x_{i}(t)}{1-x_{i}(t)}] \nonumber \\
-(1-k) \epsilon \sum_{i}^{n} \Delta x_{i} ln \frac{x_{i}(t+1)}{1-x_{i}(t+1)} 
  + (1-k) \epsilon \sum_{i}^{n} \int_{x_{i}(t)}^{x_{i}(t+1)} ln \frac{x}{1-x} dx  \label{e870a} \\
\leq -\frac{1}{2}\sum^{n}_{ij} \omega_{ij} \Delta x_{i} \Delta x_{j} 
-(1-k) \epsilon \sum_{i}^{n} [ \Delta x_{i} ln \frac{x_{i}(t+1)}{1-x_{i}(t+1)}   - \int_{x_{i}(t)}^{x_{i}(t+1)} ln \frac{x}{1-x} dx ]  \label{e871a}
\end{eqnarray}

where, $\Delta x_{i}=x_{i}(t+1)-x_{i}(t)$. In eqn.(\ref{e870a}), eqn.(\ref{e5}) is substituted.  Since $ ln \frac{x}{1-x} $ is a (strictly) monotonously increasing function for $1 > x > 0 $,
$-k \epsilon \Delta x_{i}[ ln \frac{x_{i}(t+1)}{1-x_{i}(t+1)} - ln \frac{x_{i}(t)}{1-x_{i}(t)}] \leq 0$ which is used to derive  eqn.(\ref{e871a}).  Furthermore, we get the following equations by expanding $ \int_{0}^{x_{i}(t)} ln \frac{x}{1-x} dx$ at $x_{i}(t+1)$, 

\begin{eqnarray}
\int_{0}^{x_{i}(t)} ln \frac{x}{1-x} dx
 = \int_{0}^{x_{i}(t+1)} ln \frac{x}{1-x} dx - \Delta x_{i} ln \frac{x_{i}(t+1)}{1-x_{i}(t+1)} + \frac{ [\Delta x_{i}]^{2}}{2} \frac{1}{ \varphi_{i} (1- \varphi_{i} )} \label{e873a} \\
\geq \int_{0}^{x_{i}(t+1)} ln \frac{x}{1-x} dx - \Delta x_{i} ln \frac{x_{i}(t+1)}{1-x_{i}(t+1)} + 2 [ \Delta x_{i}]^{2} \label{e874a}
\end{eqnarray}

where, $\varphi_{i}$ is between $x_{i}(t)$ and $x_{i}(t+1)$, and $min \frac{1}{\varphi_{i} (1- \varphi_{i} )} = 4$ for $0 \leq \varphi_{i} \leq 1$ is used in eqn.(\ref{e874a}). Hence the following inequality holds. 

\begin{equation}
 \int_{x_{i}(t)}^{x_{i}(t+1)} ln \frac{x}{1-x} dx   \leq \Delta x_{i} ln \frac{x_{i}(t+1)}{1-x_{i}(t+1)} - 2  [\Delta x_{i}]^{2} \label{e87a2}
\end{equation}

Therefore for $1 \geq k \geq 0$, because of eqn.(\ref{e87a2}),

\begin{equation}
f(X(t+1))-f(X(t)) \leq 
-\frac{1}{2}\sum^{n}_{ij} [ \omega_{ij}+ 4 \delta_{ij} (1-k) \epsilon ] \Delta x_{i} \Delta x_{j} \\
 =
 -\frac{1}{2}\Delta X(t)^{T} [W+ 4(1-k) \epsilon \mbox{ \boldmath$1$} ] \Delta X(t) 
 \label{e872a}
\end{equation}

where $\Delta X(t) = X(t+1)-X(t); \delta_{ij} = 1$ for $i=j$ and $\delta_{ij} = 0 $ otherwise. 

 From the assumption of this theorem, the matrix {$W+ 4(1-k) \epsilon  \mbox{ \boldmath$1$} $} is positive definite. Hence, $f(t+1)-f(t) \leq 0$, where equality holds only for $\Delta X(t) =0$, which means that $X(t)$ has reached a fixed point of the network dynamics.
Note that since $W$ is a symmetry matrix, all the eigenvalues of $W$ are real numbers, i.e., $\lambda_{min} \in \mbox{ \boldmath$  R $}$.

{\sl For condition $2$ }: 
We use a different approximation scheme to derive condition $2$. We rearrange  eqn.(\ref{e870a}) as follows:

\begin{eqnarray}
f(X(t+1))-f(X(t)) 
=-\frac{1}{2} \sum^{n}_{ij} \omega_{ij} \Delta x_{i} \Delta x_{j} - \epsilon \sum_{i}^{n} \Delta x_{i} [ ln \frac{x_{i}(t+1)}{1-x_{i}(t+1)} - ln \frac{x_{i}(t)}{1-x_{i}(t)}] \nonumber \\
-(1-k) \epsilon \sum_{i}^{n} \Delta x_{i} ln \frac{x_{i}(t)}{1-x_{i}(t)} 
  + (1-k) \epsilon \sum_{i}^{n} \int_{x_{i}(t)}^{x_{i}(t+1)} ln \frac{x}{1-x} dx  \label{e880a} \\
=-\frac{1}{2} \sum^{n}_{ij} \omega_{ij} \Delta x_{i} \Delta x_{j} - \epsilon \sum_{i}^{n} \Delta x_{i} [ ln \frac{x_{i}(t+1)}{1-x_{i}(t+1)} - ln \frac{x_{i}(t)}{1-x_{i}(t)}] \nonumber \\
+(1-k) \epsilon \sum_{i}^{n} \Delta x_{i} [ ln \frac{ \xi_{i}}{1- \xi_{i}} 
  - ln \frac{x_{i}(t)}{1- x_{i}(t)} ]  \label{e881a} 
\end{eqnarray}

where, $\xi_{i}=x_{i}(t)+ \theta_{i} \Delta x_{i}; 1 \geq \theta_{i} \geq 0$, and $\Delta x_{i} = x_{i}(t+1) - x_{i}(t)$. The mean value theorem is applied in eqn.(\ref{e881a}).

Since $ ln \frac{x}{1-x} $ is a monotonously increasing function for $1 > x > 0 $ and $\xi_{i}$ is between $x_{i}(t)$ and $ x_{i}(t+1)$,
$ ln \frac{x_{i}(t+1)}{1-x_{i}(t+1)} - ln \frac{x_{i}(t)}{1-x_{i}(t)}$ and $ ln \frac{\xi_{i}}{1-\xi_{i}}- ln \frac{x_{i}(t)}{1-x_{i}(t)}$ take the same sign. Furthermore, 

\begin{equation}
 | ln \frac{x_{i}(t+1)}{1-x_{i}(t+1)} - ln \frac{x_{i}(t)}{1-x_{i}(t)}| \geq | ln \frac{\xi_{i}}{1-\xi_{i}}-  ln \frac{x_{i}(t)}{1-x_{i}(t)}| ; (i=1,...,n)
\label{e892a}
\end{equation}

Therefore for $1 \geq k \geq 0$, because of eqn.(\ref{e892a}) and the monotonously increasing function $\ln \frac{x}{1-x}$,

\begin{eqnarray}
f(X(t+1))-f(X(t)) \leq 
-\frac{1}{2} \sum_{ij}^{n} \omega_{ij} \Delta x_{i} \Delta x_{j} 
-\sum_{i}^{n} \Delta x_{i} k \epsilon [ln \frac{x_{i}(t+1)}{1-x_{i}(t+1)} - ln \frac{x_{i}(t)}{1-x_{i}(t)}] 
 \label{e893a} \\
= -\frac{1}{2} \sum_{ij}^{n} \omega_{ij} \Delta x_{i} \Delta x_{j} 
-\sum_{i}^{n} [ \Delta x_{i}]^{2} \frac{k \epsilon }{\eta_{i}(1-\eta_{i})} \label{e894a} \\
\leq -\frac{1}{2} \Delta X(t)^{T}[W+8k \epsilon \mbox{ \boldmath$1$} ] \Delta X(t) \leq 0 \label{e895a} 
\end{eqnarray}

where $\eta_{i}$ is between $x_{i}(t)$ and $x_{i}(t+1)$. In eqn.(\ref{e893a}),eqn.(\ref{e892a}) and condition $2$ are used. The mean value theorem is applied in eqn.(\ref{e894a}) again, and $min  \frac{1}{\eta_{i} (1-\eta_{i} )} = 4$ for $0 < \eta_{i} < 1$ is substituted in eqn.(\ref{e895a}).  The equality holds only for $\Delta X(t) =0$, which means that $X(t)$ has reached a fixed point of the  network dynamics.

{\sl For condition $3$}: From eqn.(\ref{e881a}),

\begin{eqnarray}
f(X(t+1))-f(X(t)) \leq 
-\frac{1}{2} \sum_{ij}^{n} \omega_{ij} \Delta x_{i} \Delta x_{j}
-\epsilon \sum_{i}^{n} \Delta x_{i}  [ln \frac{x_{i}(t+1)}{1-x_{i}(t+1)} - ln \frac{x_{i}(t)}{1-x_{i}(t)}] 
 \label{e896a} \\
= -\frac{1}{2} \sum_{ij}^{n} \omega_{ij} \Delta x_{i} \Delta x_{j} 
-\sum_{i}^{n} [ \Delta x_{i}]^{2} \frac{ \epsilon }{\eta_{i}(1-\eta_{i})} 
\leq -\frac{1}{2} \Delta X(t)^{T}[W+8 \epsilon \mbox{ \boldmath$1$}] \Delta X(t) \leq 0 \label{e898a} 
\end{eqnarray}

where in eqn.(\ref{e896a}), eqn.(\ref{e892a}) and $k>1$ are used.  The equality holds only for $\Delta X(t) =0$.

Therefore, $f(X)$ is a Lyapunov function for eqn.(\ref{e5a}), which indicates that $X(t)$ of the system  asymptotically converges to a fixed point. 
   $\Box$ 

Marcus and Westervelt \cite{Marcus} derived the condition $1$ for $k=0$.  Therefore, Theorem \ref{t31aa} can be viewed as a generalization of the Marcus and Westervelt's result.

\newtheorem{remark}{Remark}[section]
\begin{remark}
Assume $W=W^{T}$.
According to the Gerschgorin Theorem (see Appendix \ref{appb}), 
\begin{enumerate}
\item if  $0 \leq k \leq 1/3$; $ \omega_{ii} + 4(1-k) \epsilon > |\sum_{j \neq i}^{n} \omega_{ij}|$ for $i=1,...,n$, then the matrix {$ W+ 4(1-k) \epsilon \mbox{ \boldmath$1$} $} is positive definite;
\item  if $1/3 \leq k \leq 1$; $ \omega_{ii} + 8k \epsilon > |\sum_{j \neq i}^{n} \omega_{ij}|$ for $i=1,...,n$, then the matrix {$ W+ 8k \epsilon \mbox{ \boldmath$1$}$} is positive definite;
 \item if  $1 < k $; $ \omega_{ii} + 8 \epsilon > |\sum_{j \neq i}^{n} \omega_{ij}|$ for $i=1,...,n$, then the matrix {$ W+ 8 \epsilon \mbox{ \boldmath$1$}$} is positive definite.
\end{enumerate}
 These show that the system of $X(t)$ is asymptotically stable. 
\label{rm33a}
\end{remark}

\begin{theorem}
\label{t31a}
Assume $W=W^{T}$, and
\begin{enumerate}
\item  $1/3 \geq k \geq 0$,  $4(1-k) \epsilon >- \ min_{i=1}^{n} \omega_{ii} $\\
or
\item  $1 \geq k \geq 1/3$, $8k \epsilon >- \ min_{i=1}^{n} \omega_{ii} $\\
or
\item  $ k > 1$, $8 \epsilon >- \ min_{i=1}^{n} \omega_{ii} $.
\end{enumerate}
   Then  $X(t)$ of the discrete-time system eqn.(\ref{e5a}) asymptotically converges to a fixed point, as far as eqn.(\ref{e5a}) is asynchronously updated.
\end{theorem}

{\bf Proof of Theorem \ref{t31a} }. 
Without loss of generality, let neuron-$l$ be the updated neuron at iteration-$(t+1)$, then the change in $f$ defined by eqn.(\ref{ely}) due to this updating is,

\begin{eqnarray}
f(x_{l}(t+1))-f(x_{l}(t))
=-\frac{1}{2} \omega_{ll} [ \Delta x_{l}]^{2} -k \epsilon \Delta x_{l} [ ln \frac{x_{l}(t+1)}{1-x_{l}(t+1)} - ln \frac{x_{l}(t)}{1-x_{l}(t)}] \nonumber \\
-(1-k) \epsilon \Delta x_{l} ln \frac{x_{l}(t+1)}{1-x_{l}(t+1)} 
  + (1-k) \epsilon \int_{x_{l}(t)}^{x_{l}(t+1)} ln \frac{x}{1-x} dx
\end{eqnarray}

Then, as the same way as the proof of Theorem \ref{t31aa}, we can prove this theorem by showing that $f(X)$ is also a Lyapunov function of eqn.(\ref{e5a}) with an asynchronously updating scheme.
   $\Box$

\begin{remark}
Assume $W=W^{T}$.
 If $k= 0$, the attractors of eqn.(\ref{e5a}) for both synchronously and asynchronously updating, are only fixed points and 2-periodic points (Marcus \& Westervelt, 1989) regardless of any value of $\omega_{ij}$ \ $(i,j=1,...,n)$.    
\label{rm34a}
\end{remark}

This fact can be easily shown by use of the results of Marcus and Westervelt (1989).

A state $X_{s}$ is said to be a {\bf stationary point} of $f(X)$ if satisfying $\frac{\partial f(X_{s})}{\partial X} = 0$. Therefore, stationary points are composed of {\bf local minima} where $f_{XX}$ is positive (or semi-positive) definite, {\bf local maxima} where $f_{XX}$ is negative (or semi-negative) definite, and {\bf saddle points} where $f_{XX}$ is indefinite. It should be noticed that local minima for $f(X)$ subject to $E_{0} \leq X \leq E_{1}$ include the local minima on stationary points of $f(X)$, and local minima on the boundary points or {\bf vertices} of $X$ which may not belong to stationary points of $f(X)$.

\newtheorem{lemma}{Lemma}[section]
\begin{lemma}
\label{l31aa}
Assume $W=W^{T}$.

\begin{enumerate}
\item Fixed points of eqn.(\ref{e5a}) are composed of all stationary points of $f(X)$ (i.e., local minima, local maxima and saddle points of $f(X(t))$) and all of the vertices of $X(t)$, regardless of synchronous or asynchronous updating. 
\item As far as $k \neq 1$, there is no stationary points of $f(X)$ on the vertices or boundary points of $X$.
\end{enumerate}
\end{lemma}

{\bf Proof of Lemma \ref{l31aa} }. 

 From the definition of $x_{i}(y_{i}(t))$ of eqn.(\ref{e5}) and the mean value theorem, 

\begin{equation}
x_{i}(y_{i}(t+1))-x_{i}(y_{i}(t))= [y_{i}(t+1)-y_{i}(t)] \frac{ \rho_{i}(1-\rho_{i} ) }{ \epsilon} = - \frac{\partial f(X(t))}{\partial x_{i}(t)} \frac{ \rho_{i}(1-\rho_{i} ) }{ \epsilon} 
\label{ea1}
\end{equation}
where, $\rho_{i}$ is between $x_{i}(t)$ and $x_{i}(t+1)$. In eqn.(\ref{ea1}), eqns.(\ref{e2}) and (\ref{ely}) are used.

{\sl For Lemma \ref{l31aa}-1 :} Since $E_{0} \leq X(t) \leq E_{1}$, from eqn.(\ref{ea1}), fixed points of $X(t)$ obviously satisfy 

\[ \frac{\partial f(X(t))}{\partial x_{i}(t) } = 0 \ \ \mbox{for} \ i=1,...,n \]
or

\[ x_{i}(t) =0 \ \mbox{or} \ 1 \ \ \mbox{for} \ i=1,...,n. \]
 
Therefore, the proof of Lemma \ref{l31aa}-$1$ is straightforward.

{\sl For Lemma \ref{l31aa}-2 :} The stationary points satisfy

\begin{eqnarray}
 \frac{\partial f(X(t))}{\partial x_{i}(t)} = 
-(k-1) \epsilon ln \frac{x_{i}(t)}{1-x_{i}(t)} - \sum_{j=1}^{n} \omega_{ij}x_{j}(t) -( a_{i}- \omega_{ii} a_{0i}) = 0 \ ; \ i=1,...,n. \label{esp}
\end{eqnarray}

If $k \neq 1$,  while the second term and the third term on the right  hand side in eqn.(\ref{esp}) are bounded, the first term becomes unbounded when $x_{i} \rightarrow 0$ or $1$. This fact proves the Lemma \ref{l31aa}-$2$.
 $\Box$

\begin{theorem}
\label{t32a1a}
Assume the same conditions as Theorem \ref{t31aa} or Theorem \ref{t31a}.
Then a point  $X(t)$  is  asymptotically stable for eqn.(\ref{e5a}) if and only if this point is a local minimum of $f(X)$ for $E_{0} \leq X \leq E_{1}$.\end{theorem}

{\bf Proof of Theorem \ref{t32a1a}. }  
Assume that there exists a  point $X^{*}$ which is an asymptotically stable point but not a local minimum of $f(X)$. Then, according to the definition of asymptotical stability for discrete-time systems in the sense of Lyapunov \cite{Tong}, for an arbitrary neighbourhood $U$ around $X^{*}$, there exists $V \ \subset U$ such that any trajectory with $X(0) \ \in V$ converges to $X^{*}$ with $t \rightarrow \infty$.  However, since  $X^{*}$ is not a local minimum of $f(X)$, there exists $\bar{X} \ \in V$ satisfying $f( \bar{X}) < f(X^{*})$. Therefore, when the trajectory starts from $\bar{X}$ at $t=0$, 

\begin{equation}
f(X(0))=f(\bar{X}) < f(X^{*}) = f( \lim_{t \rightarrow \infty}X(t)) = \lim_{t \rightarrow \infty} f(X(t)),
\end{equation}
which implies that the Lyapunov function increases with $t$. This contradicts Theorem \ref{t31aa} or Theorem \ref{t31a}.
Furthermore, according to Lemma \ref{l31aa} any local minimum of $f(X)$ is a fixed point of $X(t)$ for eqn.(\ref{e5a}). Therefore,  any asymptotically stable point must be a local minimum of $f(X)$.

  On the other hand, since any local minimum of $f(X)$ for $E_{0} \leq X(t) \leq E_{1}$ is a fixed point, there exists a neighbourhood $U$ around this local minimum $X^{*}$ such that  $f(X^{*})-\bar{f} =0$ and $f(X) -\bar{f} > 0$ for $X \in U$ and $X \neq X^{*}$, where $\bar{f}$ is a bounded constant.
Moreover, from Theorem \ref{t31aa} or Theorem \ref{t31a}, $f(X(t+1)) -f(X(t)) \leq 0$. Therefore any local minimum is asymptotically stable.  
  $\Box$

\newtheorem{corollary}{Corollary}[section]
 
\begin{corollary}
\label{r33a1}
Assume the same conditions of Theorem \ref{t31aa} or Theorem \ref{t31a} and assume  $\omega_{ii}=0$ for $i=1,...,n $ and $k=1$,    then a point  $X(t)$ is  asymptotically stable for eqn.(\ref{e5a}) if and only if this point is a local minimum of $f(X)$ for $E_{0} \leq X \leq E_{1}$ on the vertices.
\end{corollary}

{\bf Proof of Corollary \ref{r33a1} }. 

Since $k=1$ and $\omega_{ii}=0$, $f(X)$ defined in eqn.(\ref{ely}) is a multilinear form which has no local minimum belonging to stationary points. In other words, $f(X)$ has no local minimum for $E_{0} \leq X \leq E_{1}$ except some vertices. Therefore, we get this Corollary from Theorem \ref{t32a1a}. 
 $\Box$

Although Theorem \ref{t31aa} and Theorem \ref{t31a} are simple and easy to apply, they are  conservative estimators for the evaluation of asymptotical stability. 
Next, we give two more general theorems to evaluate the local stability in an exact manner.

\newfont{\bg}{cmr10 scaled\magstep4}
\newcommand{\bigzerol}{\smash{\hbox{\bg 0}}}
\newcommand{\bigzerou}{
  \smash{\lower1.7ex\hbox{\bg 0}}}

For synchronous updating, define the Jacobian matrix $J_{X}$ of eqn.(\ref{e5a}) at $X(t)$,

\begin{eqnarray}
 J_{X} = \frac{dX(t+1)}{dX(t)} 
 = k \left[
\begin{array}{ccc}
\frac{x_{1}(t+1)[1-x_{1}(t+1)]}{x_{1}(t)[1-x_{1}(t)]} &   & \bigzerou \\
 &   \ddots &                 \\
 \bigzerol &     & \frac{x_{n}(t+1)[1-x_{n}(t+1)]}{x_{n}(t)[1-x_{n}(t)]}  \\
\end{array}
\right]    \nonumber \\
 + \frac{1}{\epsilon}
 \left[
\begin{array}{ccc}
x_{1}(t+1)[1-x_{1}(t+1)] &   &  \bigzerou \\
 &   \ddots &                 \\
 \bigzerol &          & x_{n}(t+1)[1-x_{n}(t+1)]  \\
\end{array}
\right] W.
\label{e32a55}
\end{eqnarray}

 If $X=(x_{1},...,x_{n})$ is an internal fixed point, which means  $E_{0} < X < E_{1}, X(t)=X(t+1)=X$, then 
eqn.(\ref{e32a55}) simply becomes,

\begin{equation}
 J_{X} = k { \mbox{ \boldmath$1$}} + \frac{1}{ \epsilon} diag(x_{1}(1-x_{1}),...,x_{n}(1-x_{n})) W, 
\label{eqj}
\end{equation}

or is expressed in terms of $f_{XX}$,

\begin{equation}
 J_{X} =  { \mbox{ \boldmath$1$}} - \frac{1}{ \epsilon} diag(x_{1}(1-x_{1}),...,x_{n}(1-x_{n})) f_{XX}(X), 
\label{e32a7}
\end{equation}

where 

\begin{equation}
 f_{XX}(X) =-W- (k-1) \epsilon diag[ \frac{1}{x_{1}(1-x_{1})},..., \frac{1}{x_{n}(1-x_{n})} ].
\label{efxx} 
\end{equation}

On the other hand, if $X$ is a fixed point on boundary or vertice, for any $x_{i}=0$, 

\begin{eqnarray} 
\lim_{x_{i}(t) \rightarrow 0} \frac{x_{i}(t+1)}{x_{i}(t)} = 
\lim_{x_{i}(t) \rightarrow 0} \frac{1}{x_{i}(t)[1+e^{-\frac{1}{\epsilon}[k \epsilon ln \frac{x_{i}(t)}{1-x_{i}(t)} 
 + \sum_{j=1}^{n} \omega_{ij}x_{j}(t) + a_{i}- \omega_{ii} a_{0i} ]}]}  \\
=
\lim_{x_{i}(t) \rightarrow 0} \frac{1}{x_{i}(t)[ 1+\frac{(1-x_{i}(t))^{k}}{x_{i}(t)^{k}} e^{-\frac{1}{\epsilon}[\sum_{j=1}^{n} \omega_{ij}x_{j}(t) + a_{i}- \omega_{ii} a_{0i} ]}  ] }  \nonumber \\ =
\lim_{x_{i}(t) \rightarrow 0} x_{i}(t)^{k-1} e^{ \frac{1}{\epsilon}[\sum_{j=1}^{n} \omega_{ij}x_{j}(t) + a_{i}- \omega_{ii} a_{0i} ]}  
\end{eqnarray}

Therefore, 

\begin{eqnarray} 
\lim_{x_{i}(t) \rightarrow 0} \frac{x_{i}(t+1)}{x_{i}(t)} = 
 \left\{
\begin{array}{cc}
\infty  &  \mbox{if} \ k < 1 \\
e^{ \frac{1}{\epsilon}[\sum_{j=1}^{n} \omega_{ij}x_{j}(t) + a_{i}- \omega_{ii} a_{0i} ]}     &  \mbox{if} \ k = 1 \\
0       &  \mbox{if} \ k > 1 \\
\end{array}
\right.
\label{er1}
\end{eqnarray}

As the same way, for any $x_{i}=1$ we have

\begin{eqnarray} 
\lim_{x_{i}(t) \rightarrow 1} \frac{1-x_{i}(t+1)}{1-x_{i}(t)} = 
 \left\{
\begin{array}{cc}
\infty  &  \mbox{if} \ k < 1 \\
e^{- \frac{1}{\epsilon}[\sum_{j=1}^{n} \omega_{ij}x_{j}(t) + a_{i}- \omega_{ii} a_{0i} ]}    &  \mbox{if} \ k = 1 \\
0       &  \mbox{if} \ k > 1 \\
\end{array}
\right.
\label{er2}
\end{eqnarray}

Therefore, substituting these relations (\ref{er1})-(\ref{er2}) into eqn.(\ref{e32a55}), combining with the Jacobian matrix $J_{X}$ eqn.(\ref{eqj}) of internal point, we have the generalized Jacoban matrix of a fixed point $X^{*}=(x_{1}^{*},...,x_{n}^{*})$ for internal, boundary or vertice points,

\begin{equation}
 J_{X} = diag(k_{1},...,k_{n}) + \frac{1}{ \epsilon} diag(x_{1}^{*}(1-x_{1}^{*}),...,x_{n}^{*}(1-x_{n}^{*})) W 
\label{eqj2}
\end{equation}

where 

\begin{eqnarray}
 k_{i} = 
 \left\{
\begin{array}{cc}
\infty  & \mbox{if} \ k<1, x_{i}^{*}=0 \ \mbox{or} \ 1 \\
e^{ \frac{1-2x_{i}^{*}}{\epsilon}[\sum_{j=1}^{n} \omega_{ij}x_{j}^{*} + a_{i}- \omega_{ii} a_{0i} ]}     &  \mbox{if} \ k = 1, x_{i}^{*}=0 \ \mbox{or} \ 1 \\
0       &  \mbox{if} \ k > 1, x_{i}^{*}=0 \ \mbox{or} \ 1 \\
k       &  \mbox{if} \ x_{i}^{*} \neq 0 \ \mbox{or} \ 1 \\
\end{array}
\right\} . \label{eqk}
\end{eqnarray}

\begin{theorem}
\label{t32a11}
Assume $W=W^{T}$.

\begin{enumerate}
\item For synchronous updating, a fixed point $X^{*}$ in the region $S$ is asymptotically stable if the largest eigenvalue $ \lambda_{max}^{'} $ and smallest eigenvalue $ \lambda_{min}^{'} $ of the  matrix
 \ $ J_{X}$ \ satisfy \ \ \ 
 $ \lambda_{max}^{'}  <1$ \ and \ $ \lambda_{min}^{'} > - 1$; \  \
 on the other hand, a fixed point $X^{*}$ in the region $S$ is unstable if either  \ 
 $ \lambda_{max}^{'}  >1$ \ or \ $ \lambda_{min}^{'} <- 1$ \ is satisfied.
 $J_{X}$ is defined in eqn.(\ref{eqj2}). 

\item  For asynchronous updating,  a fixed point $X^{*}$ in the region $S$ is asymptotically stable if  \ 
$ 1 > \ max_{i=1}^{n} [k_{i} + \frac{\omega_{ii} }{\epsilon } x_{i}^{*}(1-x_{i}^{*})] $ \ and \ $ min_{i=1}^{n} [k_{i} + \frac{\omega_{ii} }{\epsilon } x_{i}^{*}(1-x_{i}^{*})] > -1$.
\end{enumerate}
\end{theorem}

{\bf Proof of Theorem \ref{t32a11} }.

{\sl For synchronous updating:}
According to dynamical systems theory \cite{Hirsch},
the fixed point  $X^{*}$ is asymptotically stable if absolute values of all eigenvalues for $J_{X}$ at $X^{*}$ are less than  $1$. On the other hand, if there exists an eigenvalue with the absolute value more than  $1$, $X^{*}$ is unstable. 
 Note that all eigenvalues of $J_{X}$ are real numbers according to Theorem \ref{appt1} of Appendix \ref{appc}.
 
{\sl For asynchronous updating}:

Without loss of generality, let neuron-$l$ be the updated neuron at iteration-(t+1), then 

 \begin{equation}
\frac{dx_{l}(t+1)}{dx_{l}(t)} = k
 \frac{x_{l}(t+1)[1-x_{l}(t+1)]}{x_{l}(t)(1-x_{l}(t))} 
 + 
\frac{ \omega_{ll} }{\epsilon}
 x_{l}(t+1)[1-x_{l}(t+1)] 
\label{e32a12}
\end{equation}

Analogously to the proof of synchronous updating, we obtain condition $2$ for asynchronous updating, by using relations (\ref{er1})-(\ref{er2}) and eqn.(\ref{e32a12}). 
  $\Box$

Let  $U_{a}$,  $D_{a}$ and  $L_{a}$  be an upper triangular, a diagonal and a lower triangular parts of $J_{X}$ defined in eqn.(\ref{eqj2}), i.e., $J_{X} = U_{a} + D_{a} + L_{a}$.
Define $\lambda_{cmin}, \lambda_{cmax}$ to be the minimal and maximal eigenvalues of the matrix $({ \mbox{ \boldmath$1$}}-L_{a})^{-1}(U_{a} + D_{a})$, respectively. Then, let us show a stricter theorem for cyclic updating.

\begin{theorem}
\label{t32a101}
Assume $W=W^{T}$. For cyclic updating,

\begin{enumerate}
\item  a fixed point $X^{*}$ in the region $S$ is asymptotically stable if the largest eigenvalue $ \lambda_{cmax} $ and smallest eigenvalue $ \lambda_{cmin}$ of the  matrix
$({ \mbox{ \boldmath$1$}}- L_{a})^{-1} ( U_{a} + D_{a})$ \ satisfy \ \ \ 
 $| \lambda_{cmax}| <1$ \ and \ $| \lambda_{cmin} |< 1$;
\item  on the other hand, a fixed point $X^{*}$ in the region $S$ is unstable if either  \ 
 $| \lambda_{cmax} | >1$ \ or \ $| \lambda_{cmin} |> 1$ is satisfied;
\item when $k < 1$, no point on vertices or boundary points is asymptotically stable.  
\end{enumerate}
\end{theorem}

{\bf Proof of Theorem \ref{t32a101} }. 

Since the cyclic updating is proceeded neuron by neuron, the total differential is, 

\begin{equation}
dx_{i}(t+1) = \sum_{j=1}^{i-1} m_{ij} dx_{j}(t+1) 
            + \sum_{j=i}^{n} m_{ij} dx_{j}(t) \ \ \ \ \mbox{for} \ i=1,...,n 
\label{e32a102}
\end{equation}

where, $m_{ij}$ is the $(i,j)$ element of the matrix $J_{X}$ defined in eqn.(\ref{eqj2}). 
 
Let us change the Jacobian matrix of cyclic updating into that of the effective synchronous updating. By dividing $J_{X}$ into an upper triangular $U_{a}$, a diagonal $D_{a}$ and a lower triangular $L_{a}$ \cite{Peterson93}, eqn.(\ref{e32a102}) can be rewritten in terms of matrix form;

\begin{equation}
dX(t+1) = L_{a}  dX(t+1) 
            + (D_{a}+U_{a}) dX(t) 
\label{e32a103}
\end{equation}

Therefore, we have the effective synchronous updating Jacobian matrix, 

\begin{equation}
\bar{J}_{X} = ({ \mbox{ \boldmath$1$}}- L_{a})^{-1}(D_{a}+U_{a}) 
\label{e32a104}
\end{equation}

Analogously to the proof of conditions $1$ and $2$ of Theorem \ref{t32a11}, the conditions $1,2$ and $3$ of this theorem can be readily proven.
  $\Box$

\begin{remark}
\label{r32a111}
Assume $W=W^{T}$,
 then in the internal points of $X$, only local minima of $f(X)$  can possibly be asymptotically stable for eqn.(\ref{e5a}), regardless of synchronous or asynchronous updating.
\end{remark}

{\bf Proof of Remark \ref{r32a111} }. 

The internal fixed points of $f(X)$ are composed of local minima, saddle points and local maxima according to Lemma \ref{l31aa}. If a fixed point $X^{*}$ is a saddle point or a local maximum, which means that $f_{XX}(X^{*})$ is indefinite or negative definite (or semi-negative definite), then at least one  eigenvalue for $diag(x_{1}^{*}(1-x_{1}^{*}),...,x_{n}^{*}(1-x_{n}^{*})) f_{XX}(X^{*})$ is negative according to Theorem \ref{appt1} of  Appendix \ref{appc}.  Hence from eqn.(\ref{e32a7}), there is at least one eigenvalue of $J_{X}$ whose absolute value is more than $1$, which means that $X^{*}$ is unstable.  Therefore, only local minimum of $f(X)$ can possibly be asymptotically stable. 
  $\Box$

According to eqn.(\ref{eqj2}), the closer $x_{i}^{*} \ (i=1,...,n)$ are  to $0$ and $1$, the smaller the absolute values of eigenvalues $\lambda_{min}^{'}$ and $\lambda_{max}^{'}$ are. Therefore, this implies that the system is likely to asymptotically converge to local minima near the vertices for $-1 <k < 1$.

\begin{remark} 
Assume $W=W^{T}$.

 \begin{enumerate}
\item When $k < 1$, no point on vertices or boundary points is asymptotically stable for both asynchronous and synchronous updating.  
\item When $k > 1$, all vertices are asymptotically stable for both synchronous and asynchronous updating. 
 \item For both synchronous updating and cyclic updating, when $k=1$, a vertex $X^{*}$ is asymptotically stable if \ \ \   $max_{i=1}^{n} e^{ \frac{1-2x_{i}^{*}}{\epsilon}[\sum_{j=1}^{n} \omega_{ij}x_{j}^{*} + a_{i}- \omega_{ii} a_{0i} ]}   <1$; \ \ 
 on the other hand, a vertex is unstable if \ 
  $max_{i=1}^{n} e^{ \frac{1-2x_{i}^{*}}{\epsilon}[\sum_{j=1}^{n} \omega_{ij}x_{j}^{*} + a_{i}- \omega_{ii} a_{0i} ]}   >1$. 
 \item For random-order updating, when $k=1$, a vertex is asymptotically stable if  \

    $max_{i=1}^{n} e^{ \frac{1-2x_{i}^{*}}{\epsilon}[\sum_{j=1}^{n} \omega_{ij}x_{j}^{*} + a_{i}- \omega_{ii} a_{0i} ]}   <1$.
 
 \end{enumerate}
\end{remark}

 From conditions $1$ and $2$ of Theorem \ref{t32a11}, Theorem \ref{t32a101} and $x_{i}^{*}= 0$ or $1$ for $i=1,...,n$, the proof of this remark is trivial.

Theorems \ref{t32a11}-\ref{t32a101} are useful for both designs of neural networks and applications of CNN and TCNN to practical problems, e.g., optimization or associative memeory, because they indicate which local minimum of the Lyapunov function is an attractor and which one is not.

\subsection{Local Bifurcations in TCNN}

Next, let us examine the bifurcations of the fixed points, in particular of 
the stationary points of $f(X)$, by taking $P$ as bifurcation parameters. $P$ is a parametric column vector and may be defined by 
 $P =( \omega_{11},..., \omega_{nn} )^{T}$ or other parameters in eqn.(\ref{e5}).

Since many applications take minima of $f(X)$ as targets of information processing, we focus on analysing
bifurcations for the stationary points of $f(X)$, especially for internal fixed points. Instead of eqn.(\ref{e5a}), we use eqn.(\ref{e5}) in the following analysis of the codimension one bifurcations of internal points, by applying the center manifold method \cite{Golubitsky,Guckenheimer}.  

Eqn.(\ref{e5}) is rewritten as, 

\begin{equation}
Y(t+1) = Y(t) - \frac{\partial f(X(Y(t)); P)}{\partial X} = Y(t) - f_{X}(X(Y(t));P) ,
\label{e5aa}
\end{equation}

   $f(X;P)$ is defined in eqn.(\ref{ely}) with parameters $P$ expressed explicitly, and  $f_{X} = \frac{\partial f}{\partial X}$.

Then the Jacobian matrix $J_{Y}(X(Y(t));P)= \frac{ \partial Y(t+1)}{ \partial Y(t)}$ of eqn.(\ref{e5aa}) or eqn.(\ref{e5}) at $(X(t),P)$ becomes

\begin{equation}
 J_{Y} = { \mbox{ \boldmath$1$}} -  f_{XY}(X(t);P) = { \mbox{ \boldmath$1$}} - 
 \frac{1}{ \epsilon } diag [  x_{1}(t)(1-x_{1}(t)),...,x_{n}(t)(1-x_{n}(t)) ]f_{XX}(X(t);P)
\label{emy}
 \end{equation}

where $f_{XY} =$ $\frac{\partial f_{X}}{\partial Y},f_{XX} =$ $\frac{\partial f_{X}}{\partial X} $ and $f_{XX}$ is a symmetrical matrix. Note that $J_{Y}$ of eqn.(\ref{emy}) is identical to $J_{X}$ of eqn.(\ref{e32a7}) for an internal point of $X$.

\begin{lemma}
\label{lb1}
Assume $W=W^{T}$, and take $P$ as bifurcation parameters,  then there is no 
Neimark-Sacker bifurcation for fixed points of eqn.(\ref{e5}). 
In other words, there only exist two types of generic codimention-one bifurcations, i.e.,
 fold bifurcations (saddle-node bifurcations) and flip bifurcations (period-doubling bifurcations). Moreover all fixed point bifurcations occur at the stationary points of $f(X)$.

\end{lemma}

{\bf Proof of Lemma \ref{lb1} }. 
According to Theorem \ref{appt1} of Appendix \ref{appc}, all eigenvalues of $J_{Y}$ are real numbers, which implies that there is no Neimark-Sacker bifurcation. In other words, only fold and flip bifurcations generally occur as codimension one bifurcations when one eigenvalue of $J_{Y}$ is equal to $1$ or $-1$ \cite{Kuznetsov,Pismen}. Eqn.(\ref{e5aa}) shows that all fixed point bifurcations occur at the stationary points of $f(X)$. 
  $\Box$

Let $Y^{*}$ or $X^{*}$ \ $(x^{*}_{i} = \frac{1}{1+ e ^{-y^{*}_{i}/ \epsilon}}$ for $i=1,...,n$) be an internal fixed point, which is in fact a stationary point of $f(X;P)$ for $W^{T} = W$ according to Lemma \ref{l31aa}. Let $P^{c}$ be the critical values for bifurcations.  Then $\frac{\partial f(X;P)}{\partial X}$ can be expanded in the vicinity of $(Y^{*}, P^{c})$ in Taylor series with respect to both variables $Y$ and parameters $P$, such that eqn.(\ref{e5aa}) becomes,

\begin{eqnarray}
Y(t+1) = Y(t) - [ f_{XY} \Delta Y + \frac{1}{2} f_{XYY} (\Delta Y \cdot \Delta Y) + \frac{1}{6} f_{XYYY}(\Delta Y \cdot \Delta Y \cdot \Delta Y) +... ]   \nonumber \\
- [f_{XP} \Delta P +  f_{XPY}(\Delta P \cdot \Delta Y) +... ] - ... , \label{e5aa1}
\end{eqnarray}
where, $\Delta Y = Y(t) - Y^{*}$, $\Delta P = P - P^{c}$ and $f_{XYY}(X^{*};P^{c})$ is a block matrix. 

Since $X^{*}$ or $Y^{*}$ is a fixed point and $P^{c}$ are the critical values of bifurcations,  $f_{X}(X^{*};P^{c}) = 0$ and at least one of eigenvalues for the Jacobian $J_{Y}(X^{*};P^{c})$ at $(X^{*}, P^{c})$ is $\lambda = 1$ or $\lambda = -1$. 

Let us first analyse the case of $\lambda = 1$  by using the center manifold method.  Let $U,U^{ \dag}$ be the normalized right and left eigenvectors of 
$J_{Y}$, respectively. That is, $J_{Y}(X^{*};P^{c}) U = U$; $U^{ \dag} J_{Y}(X^{*};P^{c}) = U^{ \dag}$ and $U^{ \dag}U =1$, or 

\begin{equation}
f_{XY}(X^{*};P^{c}) U = E_{0}.
\label{eb1}
\end{equation}

Note that $U^{ \dag}$ is a row vector.
Since $f_{XY}(X^{*};P^{c}) = \frac{1}{\epsilon } diag(x_{1}^{*}(1-x_{1}^{*}),...,x_{n}^{*}(1-x_{n}^{*}))f_{XX} $ and $x_{i}^{*}(1-x_{i}^{*}) > 0$ for $i=1,...,n$, eqn.(\ref{eb1}) implies that $f_{XX}(X^{*}; P^{c})$ has at least one zero eigenvalue.
 
 Then the center subspace at the vicinity of the bifurcation point can be defined by 

\begin{equation}
Y(t) = Y^{*} + a(t) U ,
\label{eb2}
\end{equation}

where $a(t)$ is a scalar and a sufficiently small real number.

Substituting eqn.(\ref{eb2}) into eqn.(\ref{e5aa1}) and then multiplying $U^{ \dag}$ to eqn.(\ref{e5aa1}), we have

\begin{eqnarray}
a(t+1) = a(t) - [ \frac{a(t)^{2}}{2} U^{ \dag} f_{XYY}( U \cdot U) + \frac{a(t)^{3}}{6} U^{ \dag} f_{XYYY}(U \cdot U \cdot U) + ... ]  \nonumber \\
- [ U^{ \dag} f_{XP} \Delta P +  a(t) U^{ \dag} f_{XPY}(\Delta P \cdot U) + ... ] - ... ,
\label{e5aa2}
\end{eqnarray}

which is a one-dimensional difference equation for variable $a(t)$ with parameters $\Delta P$. 

Therefore, according to bifurcation theory \cite{Guckenheimer,Kuznetsov}, the following theorem holds.

\begin{theorem}
\label{tb1}
Let $W^{T} = W$.
Assume that $X^{*}$ is an internal fixed point of eqn.(\ref{e5aa}) or eqn.(\ref{e5}) and  that $f_{XY}(X^{*};P^{c})$ has one zero eigenvalue with the right eigenvector $U$ and the left eigenvector $U^{ \dag}$.  Then for synchronous updating, a fold bifurcation (saddle-node bifurcation) occurs if  

\[ U^{ \dag} f_{XYY}(X^{*};P^{c})( U \cdot U) \neq 0
\]
and 
\[ U^{ \dag} f_{XP}(X^{*};P^{c})  \neq 0 . 
\]
\end{theorem}
 
Since the bifurcations in Theorem \ref{tb1} occur  when $f_{XX}(X^{*};P^{c})$ has zero eigenvalue and $f_{X}(X^{*};P^{c}) = 0$, they are actually stationary point bifurcations of $f(X;P)$.

\begin{remark}
\label{rb1}
The fold bifurcations for fixed points in Theorem \ref{tb1} generate or eliminate a pair of stationary points of $f(X;P)$ with the perturbations of parameters. 

\end{remark}

Next, let us analyse the case that the Jacobian matrix $J_{Y}$ has at least one eigenvalue $\lambda = -1$.  Let $X^{*}$ be a fixed point and $V, V^{ \dag}$ be the normalized right and left eigenvectors satisfying $J_{Y}(X^{*};P^{c}) V = -V; V^{ \dag} J_{Y}(X^{*};P^{c}) = - V^{ \dag}$ and $V^{ \dag}V=1$, or 

\begin{equation}
 f_{XY}(X^{*};P^{c}) V =2 V .
\label{eb21}
\end{equation}

In other words, $diag(x^{*}_{1}(1-x^{*}_{1}),...,x^{*}_{n}(1-x^{*}_{n}))f_{XX}(X^{*}; P^{c})$ has at least one eigenvalue $2 \epsilon \ (>0)$. By using Theorem \ref{appt1} of Appendix \ref{appc}, it is easy to show that $f_{XX}(X^{*};P^{c})$ has at least one positive eigenvalue at the critical point. 
 
 Then the center subspace at the vicinity of the bifurcation point can be defined by 

\begin{equation}
Y(t) = Y^{*} + b(t) V ,
\label{eb3}
\end{equation}

where $b(t)$ is a scalar and a sufficiently small real number.

As the same way of derivation of eqn.(\ref{e5aa2}), we have

\begin{eqnarray}
b(t+1) =-b(t) -  [ \frac{b(t)^{2}}{2} V^{ \dag} f_{XYY}( V \cdot V) + \frac{b(t)^{3}}{6} V^{ \dag} f_{XYYY} (V \cdot V \cdot V) +... ]  \nonumber \\
-[ V^{ \dag} f_{XP} \Delta P +  b(t) V^{ \dag} f_{XPY} (\Delta P \cdot V) +... ]  - ...,
\label{e5aa3}
\end{eqnarray}

which is a one-dimensional difference equation for variable $b(t)$ with parameters $\Delta P$. Note that $V^{ \dag}$ is a row vector.

Therefore, according to bifurcation theory \cite{Guckenheimer,Kuznetsov}, the following theorem holds.

\begin{theorem}
\label{tb2}
Let $W^{T} = W$. 
Assume that $X^{*}$ is an internal fixed point of eqn.(\ref{e5aa}) or eqn.(\ref{e5}) and that $ diag(x^{*}_{1}(1-x^{*}_{1}),...,x^{*}_{n}(1-x^{*}_{n}))f_{XX}(X^{*};P^{c})$ has one eigenvalue $2 \epsilon $ with the right eigenvector $V$ and the left eigenvector $V^{ \dag}$. 

Then for synchronous updating, a flip bifurcation (a period-doubling bifurcation) occurs if 

\[
[ \frac{V^{ \dag}}{2} f_{XYY}(X^{*};P^{c})(V \cdot V)]^{2} - \frac{V^{ \dag}}{6} f_{XYYY}(X^{*};P^{c}) (V \cdot V \cdot V) \neq 0
\]

and 

\[
2 V^{ \dag} f_{XPY}(X^{*};P^{c})  V - 
 [ V^{ \dag} f_{XP}(X^{*};P^{c}) ] 
 [ V^{ \dag} f_{XYY}(X^{*};P^{c}) (V \cdot V)]  \neq  0 
.  
\]

\end{theorem}
 
Since the flip bifurcations in Theorem \ref{tb2} occur  when $f_{XX}(X^{*};P^{c})$ has a positive eigenvalue and $f_{X}(X^{*};P^{c}) = 0$, 
these bifurcations only stabilize or unstabilize the local manifold along the direction of the eigenvector corresponding to the positive eigenvalue of $f_{XX}(X^{*};P^{c})$. Therefore, the flip bifurcations do not generally create or eliminate any stationary point of $f(X^{*};P)$, but may  unstabilize or stabilize local minima of $f(X;P)$.

\begin{remark}
\label{rb2}
The flip bifurcations for  fixed points in Theorem \ref{tb2}  occur at either a saddle point or a local minimum of $f(X;P)$.  
The flip bifurcations for fixed points cannot create (or eliminate) stationary points of $f(X^{*};P)$  with the perturbations of parameters although it may cause a change of stability of a fixed point together with the birth of a new period two orbit. 
\end{remark}
Since one eigenvalue of $f_{XX}$ is positive for the flip bifurcation, the flip bifurcation only occurs at a local minimum or a saddle point of $f(X;P)$.
As the fold  and flip bifurcations are generic, most bifurcations can be expected to be fold bifurcations and flip bifurcations when $\| P \|$ is reduced from a sufficiently large value to zero if we define $P =( \omega_{11},..., \omega_{nn} )^{T}$ as temperature parameters to realize annealing of eqn.(\ref{e5}) or eqn.(\ref{e5a}) \cite{Chen95a,Chen95b}.  
However, if a  dynamical system has certain symmetries, the other bifurcations can also be observed, such as  pitchfork bifurcations and others \cite{Sato}.

In this subsection, we only examine the bifurcations of synchronous updating for the sake of the simplification of analysis. However, the bifurcation conditions of asynchronous updating can be derived analogously to synchronous updating.

Obviously, all of analyses in this section are independent of the values of $I$ and $I_{0}$ in eqn.(\ref{e5a}). If $I_{0}=E_{0}$, eqn.(\ref{e5}) or eqn.(\ref{e5a}) is equivalent to chaotic neural networks (CNN)\cite{Aihara}.  Therefore, we have the following remark.

\begin{remark}
All of the theoretical results in this section hold for chaotic neural networks (CNN) \cite{Aihara}.
\end{remark}

\clearpage
\pagestyle{myheadings}
\markright{4 CHAOS IN TCNN}

\section{Chaos in Transiently Chaotic Neural Networks}

In the first of this section, one theorem for existence of a fixed point with sufficiently large absolute values of self feedback connection weights is derived. Then, we give sufficient conditions for chaos by applying Marotto's Theorem in the last of this section. 

Analysis in this section is mainly based on synchronous updating, but it is easy to apply it to asynchronous updating. In other words, all the theoretical results in this section hold for both synchronous and asynchronous updating.

Furthermore, for the sake of the simplification of analysis, assume that  all the self-feedback connection weights take the same value in this section, i.e., $\omega_{ii}= \omega \ (i=1,...,n)$. Therefore, $diag(\omega_{11},...,\omega_{nn})$ in eqn.(\ref{e5}) simply becomes $\omega { \mbox{ \boldmath$1$}}$ (or  $diag(\omega_{11},...,\omega_{nn}) I_{0} = \omega I_{0}$ in eqn.(\ref{e5})).

\subsection{Fixed point in TCNN}
Firstly, we show that eqn.(\ref{e5}) possesses a fixed point for sufficiently large $| \omega |$. 

\begin{theorem}
\label{t31}
Assume $1 > a_{0i} > 0$ for $ i=1,...,n$, then there exists a positive constant $c_{1}$ such that for any $| \omega | > c_{1}$ the discrete-time system of eqn.(\ref{e5}) has one fixed point  $Y^{*}$ which is bounded. If $| \omega |$ is sufficient large, $Y^{*}$ is a unique bounded fixed point.

\end{theorem}
 
  Before proving Theorem \ref{t31}, we present the following lemma.

\begin{lemma}
\label{l31}
Assume the same conditions as Theorem \ref{t31}. There exists a positive constant $c_{2}$ such that the following three conditions hold for all $ | \omega | > c_{2}$.
\begin{enumerate}
\item $W$ is invertible and $E_{0} < -W^{-1}(I- \omega I_{0}) < E_{1}$. 
\item $\bar{Y} (\omega ) = \{ Y | X(Y)= -W^{-1}(I- \omega I_{0}) \} $ is the unique fixed point of eqn.(\ref{e5}) when $k=1$.
\item $\bar{Y} ( \omega ) $ is bounded and further  $\bar{Y}_{\infty}= \lim_{| \omega | \rightarrow \infty} \bar{Y} ( \omega ) = \epsilon [ln \frac{a_{01}}{1-a_{01}}, ..., ln \frac{a_{0n}}{1-a_{0n}}]^{T}$. 
\end{enumerate}
\end{lemma}

{\bf Proof of Lemma \ref{l31}. }  Since for any $\omega \neq 0$

\begin{equation}
W = \omega \left[
\begin{array}{cccc}
1 & \omega_{12} / \omega & \cdots & \omega_{1n} / \omega   \\
\omega_{21} / \omega & 1 & \cdots & \omega_{2n} / \omega   \\
\vdots  & \vdots      & \vdots & \vdots                      \\
\omega_{n1} / \omega & \omega_{n2}/ \omega & \cdots & 1   
\end{array}
\right]
=  \omega W_{0} , \label{e6}
\end{equation}

then $\lim_{| \omega | \rightarrow \infty} W_{0} = { \mbox{ \boldmath$1$}}$, where ${ \mbox{ \boldmath$1$}}$ is the identity matrix. Therefore, $W$ is invertible for any sufficiently large $| \omega | $, i.e., $W^{-1} = \frac{1}{\omega} W^{-1}_{0}$. Moreover, 

\begin{equation}
\lim_{|\omega | \rightarrow \infty } -W^{-1}(I- \omega I_{0}) = I_{0}  \label{e7}  \end{equation}
  
Hence, by the assumption of this Lemma, there exist a $c_{2} \ (> 0)$ such that  $E_{0} < -W^{-1}(I- \omega I_{0}) < E_{1}$ for all $| \omega | > c_{2}$. 

Since $X(Y)$ is a monotonously increasing function for $Y$, and the value region of $X(Y)$ is  $E_{0} < X(Y) < E_{1}$ for a bounded $Y$ according to eqn.(\ref{e5}),  $\bar{Y} (\omega) = \{ Y | X(Y)= -W^{-1}(I- \omega I_{0}) \} $ is the unique fixed point of eqn.(\ref{e5}) when $k=1$ for all $| \omega | > c_{2}$, and is obviously bounded.

Finally, from eqns.(\ref{e5}) and (\ref{e7}),
 $\bar{Y}_{\infty}= \lim_{| \omega | \rightarrow \infty} \bar{Y} ( \omega) = 
 \epsilon [ln \frac{a_{01}}{1-a_{01}}, ..., ln \frac{a_{0n}}{1-a_{0n}}]^{T}$, 
which is also bounded.        
 $\Box$

By using Lemma \ref{l31}, we prove Theorem \ref{t31}.

{\bf Proof of Theorem \ref{t31}. }
Let $U=B^{0}(\bar{Y}(\omega ), \epsilon_{1}),$ where $\epsilon_{1}$ is a bounded arbitrary positive number.  Now, we  prove that there exists a $c_{1}(\epsilon_{1}) \   (>0)$ such that eqn.(\ref{e5}) has one fixed point $Y^{*} \in U$ for any $| \omega | > c_{1}(\epsilon_{1})$. For any $| \omega | > c_{2}$ ($c_{2}$ is given by Lemma \ref{l31}), let

\begin{equation}
Q(Y;\omega )=\frac{1}{\omega}[(k-1)Y + WX(Y) +I - \omega I_{0}] .
\label{e8}
\end{equation}

 From the condition {\it 2} of Lemma \ref{l31},

\begin{equation}
Q(\bar{Y};\omega )=\frac{1}{\omega}[(k-1)\bar{Y} + WX(\bar{Y}) +I - \omega I_{0}] = \frac{(k-1)\bar{Y}}{\omega } .
\label{e9}
\end{equation}

Then, according to the condition {\it 3} of Lemma \ref{l31}, 

\begin{equation}
\lim_{| \omega | \rightarrow \infty} \| Q(\bar{Y}; \omega )\| = 0,
\label{e88}
\end{equation}

 which means $Y=\bar{Y}(\omega )$ is an approximate solution of $Q(Y;\omega) =0$ for a sufficiently large number $| \omega |$.

Next, from eqn.(\ref{e8}), the Jacobian matrix is

\begin{equation}
Q_{Y}(Y;\omega )=\frac{1}{\omega}[(k-1){ \mbox{ \boldmath$1$}} + \frac{1}{\epsilon}WZ(Y) ]
                     =\frac{(k-1)}{\omega}{ \mbox{ \boldmath$1$}} + \frac{1}{\epsilon}W_{0}Z(Y)
\label{e10}
\end{equation}

, where  
\begin{equation}
Z(Y) = \left[
\begin{array}{ccc}
x_{1}(y_{1}) [1-x_{1}(y_{1})] &        &  \bigzerou              \\
               & \ddots &               \\
\bigzerol      &        & x_{n}(y_{n})[1-x_{n}(y_{n})] 
\end{array}
\right]
\label{e11}
\end{equation}

Since for any bounded $Y \in U$ including $\bar{Y}$, 

\[
\lim_{| \omega | \rightarrow \infty} det( Q_{Y}(Y;\omega ))  = \lim_{|\omega | \rightarrow \infty} det(\frac{1}{\epsilon}W_{0}Z(Y)) 
 = \lim_{|\omega | \rightarrow \infty} \frac{1}{\epsilon} \prod_{i=1}^{n}x_{i}(y_{i})[1-x_{i}(y_{i})] \neq 0 ,
\]

then,  there exists  
 a $c_{3}(\epsilon_{1}) \   (>0)$ such that $\| Q_{Y}(\bar{Y}; \omega)^{-1}\| < d_{1}$ and 
$\| Q_{Y}(Y; \omega) -Q_{Y}(\bar{Y}; \omega) \| < \mu_{1} /d_{1}$ for a sufficiently large positive number $d_{1}$ and for a positive number $\mu_{1} \ (<1)$ with any $| \omega | > c_{3}(\epsilon_{1})$ and any $Y \in U$.
   
Therefore,  there exists a $r_{1} \   (>0)$ such that $r_{1} < (1- \mu_{1}) \epsilon_{1} / d_{1}$ and,  according to eqn.(\ref{e88}), there exists $c_{4} \   (>0)$ such that $\| Q(\bar{Y}; \omega) \| < r_{1} $ for all $| \omega | > c_{4}$.  Let $c_{1}(\epsilon_{1}) = max(c_{2},c_{3}(\epsilon_{1}), c_{4})$, then the following three conditions hold for all $| \omega | > c_{1}(\epsilon_{1})$.

\begin{enumerate}
\item $ \Omega (\epsilon_{1}) =\{ Y | \| Y- \bar{Y} \| < \epsilon_{1} \} \subset U $
\item $\| Q_{Y}(Y; \omega) -Q_{Y}(\bar{Y}; \omega) \| < \mu_{1} /d_{1}$ 
\item $\frac{r_{1} d_{1}}{1- \mu_{1}} < \epsilon_{1}$, where $\| Q(\bar{Y}; \omega) \| < r_{1} $ and $\| Q_{Y}(\bar{Y}; \omega)^{-1}\| < d_{1}$.
\end{enumerate}

According to Urabe's Proposition (Urabe, 1965; see the Appendix \ref{appa})\cite{Urabe}, $Q(Y; \omega) =0$ has a unique solution $Y^{*}(\omega)$ in $\Omega (\epsilon_{1}) \subset U$ for all $| \omega | > c_{1}(\epsilon_{1})$.
That is,

\begin{equation}
\frac{1}{\omega}[(k-1)Y^{*} + WX(Y^{*}) +I - \omega I_{0}]= 0
\label{e12}
\end{equation}

or 

\begin{equation}
Y^{*}= kY^{*} + WX(Y^{*}) +I - \omega I_{0} = f(Y^{*})
\label{e13}
\end{equation}

which means that eqn.(\ref{e5}) has one unique fixed point $Y \in U$ for all $| \omega | > c_{1}(\epsilon_{1})$. Since $U$ is a bounded set, $Y^{*}$ is bounded.

Next, let us show that $Y^{*}$ is a unique fixed point for all bounded region if $| \omega |$ is sufficient large. Assume $Y^{0}$ to be another bounded fixed point different from $Y^{*}$, in other words $Y^{0} \not\in U$. Since both $Y^{*}$ and $Y^{0}$ are fixed points of eqn.(\ref{e5}) and $W$ is invertible, the following relation holds.

\[ X(Y^{*})-X(Y^{0}) = -(k-1)W^{-1}(Y^{*}-Y^{0})=-\frac{k-1}{ \omega }W_{0}^{-1}(Y^{*}-Y^{0})
\]
 
Therefore, $X(Y^{*})-X(Y^{0})$ or $Y^{*}-Y^{0}$ approaches zero when $| \omega |$ is sufficient large, because $Y^{*}$ and $Y^{0}$ are bounded. Since $Y^{*} \in U$, there exists a sufficient large $c_{0} \ (>0)$ such that $Y^{0} \in U$ for all $| \omega | > c_{0}$, which contradicts the uniqueness of the fixed point in $U$. Therefore, the theorem is proved. 
  $\Box$

\begin{remark}
Assume the same conditions as Theorem \ref{t31}.
Define $Y^{*}_{\infty} = \lim_{| \omega | \rightarrow \infty} Y^{*}$, then 
$Y^{*}_{\infty}= \bar{Y}_{\infty} = 
 \epsilon [ln \frac{a_{01}}{1-a_{01}}, ..., ln \frac{a_{0n}}{1-a_{0n}}]^{T}$, which is also a unique fixed point (bounded) for $| \omega | \rightarrow \infty$.
\label{rm31}
\end{remark}

{\bf Proof of Remark \ref{rm31} }. Since $Y^{*}_{\infty}$ is bounded, and  

\begin{eqnarray}
\lim_{| \omega | \rightarrow \infty} 
\frac{1}{\omega}[(k-1)Y^{*}_{\infty} + WX(Y^{*}_{\infty}) +I - \omega I_{0}]
= \lim_{| \omega | \rightarrow \infty} [ \frac{(k-1)Y^{*}_{\infty}+ I}{\omega}
   + W_{0}X(Y^{*}_{\infty}) - I_{0} ] \nonumber \\
=   X(Y^{*}_{\infty}) - I_{0} =0. 
\end{eqnarray}

Hence, the Remark is proven.    $\Box$

According to eqn.(\ref{efxx}), $f_{XX}$ is positive definite if $\omega$ is sufficiently small (negative). On the other side, $f_{XX}$ is negative definite if $\omega$ is sufficiently large (positive).

\begin{remark}
$Y^{*}$ or $X(Y^{*})$ is a unique minimum of $f$ defined in eqn.(\ref{ely}) if $\omega$ is sufficiently small (negative). On the other side, $Y^{*}$ is a unique maximum of $f$ if $\omega$ is sufficiently large (positive).  
\end{remark}

\begin{remark}
Assume the same conditions as Theorem \ref{t31}.
If $a_{0i} =0 \ (i=1,...,n)$, there is no bounded fixed point for sufficiently large $| \omega |$.
\label{rm32}
\end{remark}

The proof of this remark is straightforward from Remark \ref{rm31}.

\subsection{Existence of chaos in TCNN}
Next, we establish a principal result of this section.

\begin{theorem}
\label{t32}
Assume $1 > a_{0i} > 0$ for $i=1,...,n$, and 

\begin{enumerate}
\item  $k > \frac{a_{0i}}{1-a_{0i}}$ for $i=1,...,n$, $\omega <0$;\\
or 
\item  $k< -1$,  $ \omega >0$;\\
or 
\item  $ k> \frac{1-a_{0i}}{a_{0i}} $ for $i=1,...,n$, $\omega <0$,
\end{enumerate}

then there exists a positive constant $c_{5}$ such that for any $| \omega | > c_{5}$ the discrete-time system of eqn.(\ref{e5}) is chaotic in the sense of Marotto.
\end{theorem}

Theorem \ref{t32} can be proved by the following two preliminary lemmas.

\begin{lemma}
\label{l32}
There exists a positive constant $c_{6}$ such that all eigenvalues of $F_{Y}(Y)$ exceed the unity in norm for all $| \omega | > c_{6}$ and for any bounded $Y$.
\end{lemma}

{\bf Proof of Lemma \ref{l32}}. From eqn.(\ref{e5}),

\begin{equation}
F_{Y}(Y) = k{ \mbox{ \boldmath$1$}} + \frac{1}{\epsilon}WZ(Y) =
  \left[
\begin{array}{ccc}
k+\omega x_{1}(1-x_{1})/ \epsilon & \cdots & \omega_{1n}x_{n}(1-x_{n})/ \epsilon \\ 
\vdots                            & \vdots & \vdots      \\
\omega_{n1}x_{1}(1-x_{1})/ \epsilon & \cdots & k+\omega x_{n}(1-x_{n})/ \epsilon 
\end{array}
\right]
\end{equation}  

Let $\lambda_{p} ; (p=1,...,n)$ be eigenvalues of $F_{Y}(Y)$ for any bounded $Y$, and $\theta_{i} =   k+ \omega x_{i}(1-x_{i}) / \epsilon ;
r_{i} = \sum_{j \neq i}^{n} | \omega_{ij}| x_{j}(1-x_{j})/ \epsilon  ; (i=1,...,n)$.  Then according to Gerschgorin Theorem (see Appendix \ref{appb}), $\lambda_{p} \in \bigcup_{i=1}^{n} B(\theta_{i};r_{i}) ; (p=1,...,n).$ In other words, 
there exists a set $S_{0} \subset \{1,...,n \}$ such that the following inequalities hold  for some $i \in S_{0}$.

\begin{equation}
r_{i} \geq | \theta_{i} - \lambda_{p} | \geq  | \theta_{i} | - | \lambda_{p} | 
 \geq  | \omega x_{i}(1-x_{i}) / \epsilon | -k - |\lambda_{p} | ; (p=1,...,n)
\end{equation} 

 Hence, $| \lambda_{p}| \geq  |\omega| x_{i}(1-x_{i}) / \epsilon - k 
- \sum_{j \neq i}^{n} | \omega_{ij}|x_{j}(1-x_{j})/ \epsilon $ \ for $ p=1,...,n$ and $i \in S_{0}$.

Since $Y$ is bounded, $x_{i} \neq 0$ or $1$. Define $c_{6} = max_{i=1}^{n} [(1+k)\epsilon +  \sum_{j \neq i}^{n} | \omega_{ij}|x_{j}(1-x_{j}) ]/[x_{i}(1-x_{i})]$, then for any $| \omega | > c_{6}$, $| \lambda_{p} | >1 \ (p=1,...,n)$, which proves the Lemma.     $\Box$

\begin{lemma}
\label{l33}
Assume the same conditions as Theorem \ref{t32}, then $Y^{*}$ is a snap-back repeller for any $| \omega | > c_{5}$.  
\end{lemma}

To prove Lemma \ref{l33}, we have to show that, (a) there exists a fixed point $Y^{*}$; (b) there is a bounded $r$ such that the absolute value of all eigenvalues of $F_{Y}$ for $B(Y^{*};r)$ is larger than $1$; (c) there is a point $Y^{0*} \in B(Y^{*};r)$ and $Y^{0*} \neq Y^{*}$ such that $F^{m}(Y^{0*}) = Y^{*}$ holds, when $| \omega | $ is sufficiently large. Actually, Theorem \ref{t31} and Lemma \ref{l32} ensure (a) and (b). Next, we find a $Y^{0*}$ to prove $F^{2}(Y^{0*}) = Y^{*}$.

{\bf Proof of Lemma \ref{l33}}.  We first prove the conditions $1$ and $2$ of Theorem \ref{t32}, and then show the condition $3$.

For any $| \omega | > c_{1}$, define $(H_{1},...,H_{n})$ as 
\begin{equation}
[ H_{1},...H_{n} ]^{T}= W_{0}^{-1}.
\label{e15}
\end{equation}

{\sl Proof of conditions of {\sl 1} and {\sl 2} }:

 Let $G=(g_{1},...,g_{n})^{T}$, where, 

\begin{equation}
g_{i} = \frac{H_{i}^{T}}{\omega } 
[ 
\frac{Y^{*}-I+ \omega I_{0}}{k} -I + \omega I_{0} - \epsilon k
\left[
\begin{array}{c}
ln \frac{a_{01}(1+1/k)}{1-a_{01}(1+1/k)}   \\
\vdots   \\
ln \frac{a_{0n}(1+1/k)}{1-a_{0n}(1+1/k)} 
\end{array}
\right] ]
\label{e17}
\end{equation}

Then define
\begin{equation}
 Y^{0} = \epsilon [ln \frac{g_{1}}{1-g_{1}},..., ln \frac{g_{n}}{1-g_{n}} ]^{T}. \label{e18}
\end{equation} 

Therefore, $X(Y^{0})= G$ by eqns.(\ref{e5}),(\ref{e17}) and (\ref{e18}). Note that the following condition should be satisfied  in eqn.(\ref{e18}).

\begin{equation}
0 < g_{i} < 1  \ \ (i=1,..,n)                            \label{e20} 
\end{equation}

Since  $\lim_{| \omega | \rightarrow \infty} g_{i}=
 (1+1/k)a_{0i} \ (i=1,...,n)$ because of $\lim_{| \omega | \rightarrow \infty} W_{0}^{-1} = { \mbox{ \boldmath$1$}}$, hence, according to the conditions $1$ and $2$, there exists a $c_{7} \   (>0)$ such that eqn.(\ref{e20}) holds for all $| \omega | > c_{7}$.

Since $g_{i} \ (i=1,...,n)$ is bounded, $Y^{0}$ is also bounded. Actually, from eqn.(\ref{e18})

\begin{equation}
\lim_{| \omega | \rightarrow \infty}Y^{0} = 
  \epsilon [ln \frac{a_{01}(1+1/k)}{1-a_{01}(1+1/k)},...,ln \frac{a_{0n}(1+1/k)}{1-a_{0n}(1+1/k)}]^{T}  
\end{equation}

which is not equal to $Y^{*}_{\infty}$ according to Remark \ref{rm31}. Therefore, there exists a positive numbers $c_{8}$ and $\epsilon_{2}$ such that $Y^{*} \not\in B(Y^{0}; \epsilon_{2})$ for any $| \omega |> c_{8}$.

Furthermore, from eqns.(\ref{e5}) and (\ref{e18}),

\begin{equation}
F(Y^{0}) = \frac{1}{k}(Y^{*}- I+ \omega I_{0}) +k \epsilon  
\left[
\begin{array}{c}
ln \frac{g_{1}}{1-g_{1}} -ln \frac{a_{01}(1+1/k)}{1-a_{01}(1+1/k)}   \\
\vdots   \\
ln \frac{g_{n}}{1-g_{n}} -ln \frac{a_{0n}(1+1/k)}{1-a_{0n}(1+1/k)} 
\end{array}
\right]
\label{e21}
\end{equation}

Next, we use Urabe's proposition again to prove the lemma.

Let 
\begin{equation}
Q(Y; \omega) = \frac{1}{\omega} [F(F(Y))-Y^{*} ] = \frac{1}{\omega} [F^{2}(Y)-Y^{*} ]
\label{e22}
\end{equation}

then
\begin{equation}
Q(Y^{0}; \omega) =\frac{k^{2} \epsilon}{\omega}  
\left[
\begin{array}{c}
ln \frac{g_{1}}{1-g_{1}} -ln \frac{a_{01}(1+1/k)}{1-a_{01}(1+1/k)}   \\
\vdots   \\
ln \frac{g_{n}}{1-g_{n}} -ln \frac{a_{0n}(1+1/k)}{1-a_{0n}(1+1/k)} 
\end{array}
\right] + \frac{1}{\omega}WX(F(Y^{0}))
\label{e23}
\end{equation}

 From eqns.(\ref{e20}) and (\ref{e21}), for sufficiently large $| \omega |$
\begin{equation}
F(Y^{0}) \propto  \frac{\omega I_{0}}{k}
\label{e270}
\end{equation}
 which is negative because of the conditions of $1$ and $2$. 
Hence, 

\begin{equation}
\lim_{| \omega | \rightarrow \infty} WX(F(Y^{0})) =E_{0}
\end{equation}

Therefore,

\begin{equation}
\lim_{| \omega | \rightarrow \infty} \| Q(Y^{0}; \omega) \| =0
\label{e25}
\end{equation}

 which means $Y=Y^{0}(\omega )$ is an approximate solution of $Q(Y;\omega) =0$ for a sufficiently large number $| \omega |$.

Define {$ \mbox{ \boldmath$0$} $} as a $n \times n$ zero matrix where all elements are zero.  By eqn.(\ref{e22}), 

\begin{equation}
Q_{Y}(Y;\omega )=[k{ \mbox{ \boldmath$1$}}/ \omega + W_{0}Z(Y)/ \epsilon ]
                     [k{ \mbox{ \boldmath$1$}} + \omega W_{0}Z(F(Y))/ \epsilon . ]
\label{e26}
\end{equation}

Since 
\begin{equation}
\lim_{| \omega | \rightarrow \infty} det(W_{0}Z(Y^{0})) = \prod_{i=1}^{n}
(1+1/k)a_{0i}[1-(1+1/k)a_{0i}] \neq  0 
\end{equation}

\begin{equation}
\lim_{| \omega | \rightarrow \infty} k { \mbox{ \boldmath$1$}} / \omega = \mbox{ \boldmath$0$}
\end{equation}

and

\begin{equation}
\lim_{| \omega | \rightarrow \infty}  \frac{\omega}{\epsilon}W_{0}Z(F(Y^{0}))  = \mbox{ \boldmath$0$} 
\end{equation}
because of the conditions of $1$ and $2$ and eqn.(\ref{e270}),

 $\lim_{| \omega | \rightarrow \infty} det(Q_{Y}(Y^{0}; \omega)) = \frac{k}{ \epsilon} \prod_{i=1}^{n}
(1+1/k)a_{0i}[1-(1+1/k)a_{0i}] \neq  0 $

which implies that  there exists  
a $c_{9} \  (>0)$ such that for  a sufficiently large positive number $d_{2}$,   $\| Q_{Y}(Y^{0}; \omega)^{-1}\| < d_{2}$ 
 for any $| \omega | > c_{9}$.
Furthermore, for a positive number  
$\mu_{2} \ (<1)$, there exists a positive number $\epsilon_{3} \ (< \epsilon_{2})$ such that $\| Q_{Y}(Y; \omega) -Q_{Y}(Y^{0}; \omega) \| < \mu_{2} /d_{2}$ for all $Y \in B(Y^{0}; \epsilon_{3})$

Therefore,  there exists a $r_{2} \  (>0)$ such that $r_{2} < (1- \mu_{2}) \epsilon_{2} / d_{2}$ and,  according to eqn.(\ref{e25}), there exists $c_{10}(\epsilon_{2}) \  (>0)$ such that $\| Q(Y^{0}; \omega) \| < r_{2} $ for all $| \omega | > c_{10}(\epsilon_{2})$.  Let $c_{11}(\epsilon_{2}) = max(c_{1},c_{7},c_{8},c_{9},c_{10}(\epsilon_{2}))$, then the following three conditions hold for all $| \omega | > c_{11}(\epsilon_{2})$.

\begin{enumerate}
\item $ \Omega (\epsilon_{3}) = \{ Y | \| Y- Y^{0} \| < \epsilon_{3} \} \subset B(Y^{0}; \epsilon_{2}) $
\item $\| Q_{Y}(Y; \omega) -Q_{Y}(Y^{0}; \omega) \| < \mu_{2} /d_{2}$ 
\item $\frac{r_{2} d_{2}}{1- \mu_{2}} < \epsilon_{3}$, where $\| Q(Y^{0}; \omega) \| < r_{2} $ and $\| Q_{Y}(Y^{0}; \omega)^{-1}\| < d_{2}$.
\end{enumerate}

According to Urabe's Proposition (see the Appendix \ref{appa}) \cite{Urabe}, $Q(Y; \omega) =0$ has the unique solution $Y^{0*}(\omega)$ in $\Omega (\epsilon_{3}) \subset B(Y^{0};\epsilon_{2})$ for all $| \omega | > c_{11}(\epsilon_{2})$, which also implies $Y^{0*} \neq Y^{*}$ according to the definition of $B(Y^{0}; \epsilon_{2})$.
That is,
 
\begin{equation}
Y^{*} = F(F(Y^{0*}))= F^{2}(Y^{0*})
\label{e30}
\end{equation}

for all $| \omega | > c_{11}(\epsilon_{2})$. By Lemma \ref{l32}, there exists a $c_{12}  \ (>0)$ and $\epsilon_{4}  \ (>0)$ such that all eigenvalues of $F_{Y}(Y)$ exceed the unity in norm for
all $| \omega | > c_{12}$ and for any $Y \in B(Y^{*}; \epsilon_{4})$, where $Y^{0*} \in B(Y^{*}; \epsilon_{4})$.  Let $c_{5} = max(c_{11}, c_{12})$, then $Y^{*}$ is a snap-back repeller.

{\sl Proof of condition {\sl 3}}:

 Let $G=(g_{1},...,g_{n})^{T}$, where, 

\begin{equation}
g_{i} = \frac{H_{i}^{T}}{\omega } 
[ 
\frac{Y^{*}-I - \omega (E_{1} -I_{0})}{k} -I+ \omega I_{0}  - \epsilon k
\left[
\begin{array}{c}
ln \frac{a_{01}(1+1/k)-1/k}{1-a_{01}(1+1/k) +1/k}   \\
\vdots   \\
ln \frac{a_{0n}(1+1/k)-1/k}{1-a_{0n}(1+1/k) +1/k} 
\end{array}
\right] ]
\end{equation}

Define
\begin{equation}
 Y^{0} = \epsilon [ln \frac{g_{1}}{1-g_{1}},..., ln \frac{g_{n}}{1-g_{n}}]^{T}.
\end{equation} 

Note that the following condition should be satisfied,

\begin{equation}
0 < g_{i} < 1  ;(i=1,..,n).                            \label{e201} 
\end{equation}

Furthermore,

\begin{equation}
F(Y^{0}) = \frac{1}{k}[Y^{*}- I- \omega(E_{1}- I_{0})] +k \epsilon  
\left[
\begin{array}{c}
ln \frac{g_{1}}{1-g_{1}} - ln \frac{a_{01}(1+1/k)-1/k}{1-a_{01}(1+1/k)+1/k}  \\
\vdots   \\
ln \frac{g_{n}}{1-g_{n}} - ln \frac{a_{0n}(1+1/k)-1/k}{1-a_{0n}(1+1/k)+1/k}
\end{array}
\right]
\label{e211}
\end{equation}

Hence, for sufficiently large $| \omega |$, 
\begin{equation}
F(Y^{0}) \propto - \frac{\omega (E_{1}-I_{0})}{k}
\label{e2701}
\end{equation}
 which is positive because of the condition $3$. Note that the condition $2$ also satisfies eqns.(\ref{e201}) and (\ref{e2701}). 

Then as the same way as the proof of conditions $1$ and $2$, condition $3$ can be  proved.     $\Box$

{\bf Proof of Theorem \ref{t32} }.  By Lemma \ref{l33} and Marotto's Theorem, the proof is straightforward.    $\Box$ 

Obviously, there are two different type of snap-back repellers for Lemma \ref{l33}.   One  is $ \{ Y^{0*} \Longrightarrow  \ \mbox{(sufficiently negative)} \ F(Y^{0*})  \Longrightarrow  Y^{*} \} $  corresponding to conditions $1$ and $2$, the other is $ \{ Y^{0*} \Longrightarrow  \ \mbox{(sufficiently positive)} \ F(Y^{0*}) \Longrightarrow  Y^{*} \} $ corresponding to conditions $3$ and $2$.

\begin{remark}
Chaos in eqn.(\ref{e5}) is generated not from a saddle point, but from a repeller when $| \omega |$ is sufficiently large.  
\label{rm33}
\end{remark}

\begin{theorem}
 $Y(t)$ of eqn.(\ref{e5}) is bounded for any initial value $Y(0)$ if $-1 < k < 1$.
\label{rm34}
\end{theorem}

{\bf Proof of Theorem \ref{rm34} }. Let $ d_{i} = \sum_{j=1}^{n} |\omega_{ij}| + |a_{i}- \omega_{ii}a_{0i}|,$ which is bounded. According to eqn.(\ref{e3}) 

\begin{equation}
 ky_{i}(t)-d_{i} \leq y_{i}(t+1) \leq ky_{i}(t) +d_{i}; \ (t=0,1,...) 
\label{equ1}
\end{equation}
 
{\sl For the case of $k \geq 0$ :}  From the right hand side of eqn.(\ref{equ1}) for $t,t-1,...0$,

\begin{equation}
 y_{i}(t+1) \leq k^{2}y_{i}(t-1) +kd_{i}+d_{i} \leq ... \leq k^{t+1}y_{i}(0)+ 
\frac{1-k^{t+1}}{1-k}d_{i}
\label{equ2}
\end{equation}

On the other side, from the left hand side of eqn.(\ref{equ1}) for $t,t-1,...0$,
\begin{equation}
 y_{i}(t+1) \geq k^{2}y_{i}(t-1) -kd_{i}-d_{i} \geq ... \geq k^{t+1}y_{i}(0)- 
\frac{1-k^{t+1}}{1-k}d_{i}
\label{equ3}
\end{equation}

Since $y_{i}(0), d_{i}$ are bounded and $|k^{t+1}| <1$ for $t=0,1,...$, $y_{t+1}(t+1)$ is bounded. 

{\sl For the case of $k <0$ :} As the same way, we can show 

\begin{equation}
k^{t+1}y_{i}(0)- \frac{1-(-k)^{t+1}}{1+k}d_{i} \leq 
 y_{i}(t+1) \leq k^{t+1}y_{i}(0)+ \frac{1-(-k)^{t+1}}{1+k}d_{i},
\label{equ4}
\end{equation}

 thereby proving that any $y_{i}(t+1)$ is bounded for any initial value $y_{i}(0)$.
$\Box$ 

As a matter of fact, for steady states, i.e., when $t$ is sufficiently large,
$ k^{t+1} \ \mbox{or } \ (-k)^{t+1} \longrightarrow 0 $.  That means that, $|y_{i}(t+1)| \leq \frac{d_{i}}{1-|k|}$ according to eqns.(\ref{equ2})-(\ref{equ4}) for any $y_{i}(0)$ when $t$ is sufficiently large.

\newtheorem{example}{Example}[section]
\begin{example}
\label{ex31}
 Consider a one-dimensional case of eqn.(\ref{e5}),

\[ y(t+1) = ky(t)+ \omega x(y(t)) +a- \omega a_{0} \ ; \ x(y(t))=\frac{1}{1+e^{-y(t)/ \epsilon }}. \]

Let $k=0.9; \epsilon =1/250; a=0; a_{0}=0.65$. Figure \ref{f1} shows the output of the neuron and the Lyapunov exponent \cite{Chen95b} with increasing $| \omega | $. 
\end{example}

We first use the condition $2$ of Theorem \ref{t31aa} or Theorem \ref{t31a} to estimate the stability of this example. 
 According to the condition $2$ of Theorem \ref{t31a}, the estimated value of $\omega $ is $\omega_{e}= - 8k \epsilon = - 0.0288$, below which the system can be ensured to be stable. If $- \omega < 0.0288$, the system is stable and converges to a fixed point, which is verified by the numerical simulation in Figure \ref{f1}. 

Next, we use Theorem \ref{t32a11} to calculate the first bifurcation point and to analyse the dynamics exactly. The fixed point satisfies

\begin{equation}
 (k-1) \epsilon ln \frac{x(t)}{1-x(t)}+ \omega x(t) - \omega a_{0}=0.
\label{ee1}
\end{equation}

According Theorem \ref{t32a11}, the condition under which the fixed point is stable follows

\begin{equation}
k + \frac{1}{\epsilon} \omega x(t)(1-x(t)) > -1, 
\label{ee2}
\end{equation}

because we only analyse the dynamical behaviour for $k>0$ and $\omega <0$.
Therefore, by solving eqns.(\ref{ee1}) and (\ref{ee2}), we obtain the critical values of $\omega$ and the fixed point $x^{*}$, 

\[\omega_{c} = -0.033104 ; x^{*c} = 0.64289. \]

Therefore,  If $- \omega < 0.033104$, the system is stable and converges to a fixed point; at $-\omega =0.033104$, the fixed point bifurcates into 2-periodic points which is confirmed by Theorem \ref{tb2};  when $- \omega > 0.033104$, the system becomes unstable.  These facts are verified by the numerical simulation in Figure \ref{f1}.

Finally, we use the  condition $3$ of Theorem \ref{t32} to analyze chaotic dynamics of this example. 
 According to the condition $1$ of Theorem \ref{t32}, there exists chaotic behavior when $-\omega $ is sufficiently large.  Actually, in this example, when $ - \omega > 0.0596$, the chaotic behavior appears, which is numerically verified by positive Lyapunov exponent in Figure \ref{f1}, although there also exist many periodic windows.

\clearpage
\pagestyle{myheadings}
\markright{5 ASYMPTOTICAL STABILITY OF DRNN}

\section{Asymptotical Stability in Discrete-time Recurrent Neural Networks}

In this section, we firstly give a general form of discrete-time recurrent neural networks. Then, several convergence theorems for symmetrical versions of DRNN  are addressed. All the theorems in this section can be proved by similar arguments as in section 3.

\subsection{Difference scheme}

The continuous-time recurrent neural network can generally be written in forms of differential equations as follows (Amari, 1972; Hopfield,1984):                                                
 \begin{eqnarray}
x_{i}=\frac{1}{1+e^{- y_{i} / \epsilon }}  \nonumber     \\
\frac{dy_{i}}{dt} = - \mu y_{i} + \sum_{j=1}^{n} \omega_{ij}x_{j} + a_{i} \label{e42}  \ \ 
(i=1,...,n)                                              
\end{eqnarray}

where all variables and parameters are real numbers. 

Then, the discrete-time recurrent neural networks (DRNN) on the basis of the Euler method are given in eqns.(\ref{e45}).

\begin{equation}
Y(t+1)  = F(Y(t)) \ = k Y(t) + \Delta t [ WX(Y(t)) + I] 
\label{e45}
 \end{equation}
 
where $\Delta t > 0$, $k=1- \mu \Delta t$ and, 

\[
Y(t) = \left[
\begin{array}{c}
y_{1}(t) \\ \vdots \\ y_{n}(t)
\end{array}
\right]
, \ \ \ 
X(Y(t)) = \left[
\begin{array}{c}
x_{1}(t) \\ \vdots \\ x_{n}(t)
\end{array}
\right] 
=
 \left[
\begin{array}{c}
\frac{1}{1+e^{-  y_{1}(t) / \epsilon }} \\ \vdots \\ \frac{1}{1+e^{-  y_{n}(t) / \epsilon }}
\end{array}
\right]
\]

\[
I = \left[
\begin{array}{c}
a_{1} \\ \vdots \\ a_{n}
\end{array}
\right]
, \ \ \
W = \left[
\begin{array}{cccc}
\omega_{11} & \omega_{12} & \cdots & \omega_{1n} \\
\omega_{21} & \omega_{22} & \cdots & \omega_{2n} \\
\vdots  & \vdots      & \vdots & \vdots      \\
\omega_{n1} & \omega_{n2} & \cdots & \omega_{nn} 
\end{array}
\right]
\]

Eqn.(\ref{e45}) is  a neural network model with discrete-time and continuous-state \cite{Marcus}.
Obviously, $F$ in eqn.(\ref{e45}) is a $C^{1}$ class function for any bounded $Y$.

To analyse asymptotical stability of $X(t)$, we rewrite eqn.(\ref{e45}) by explicitly expressing $X(t)$,

\begin{equation}
X(t+1) =\bar{F}(X(t)) 
=
 \left[
\begin{array}{c}
\frac{1}{1+e^{-\frac{1}{\epsilon}[k \epsilon ln \frac{x_{1}(t)}{1-x_{1}(t)} 
 + \sum_{j=1}^{n} \Delta t \omega_{1j}x_{j}(t) + \Delta t a_{1} ]}} \\
 \vdots \\ 
\frac{1}{1+e^{-\frac{1}{\epsilon}[k \epsilon ln \frac{x_{n}(t)}{1-x_{n}(t)} 
 + \sum_{j=1}^{n} \Delta t \omega_{nj}x_{j}(t) + \Delta t a_{n} ]     }}
\end{array}
\right],
\label{e45a}
\end{equation}

where, $ 0 \leq x_{i}(t) \leq 1 $ for $i=1,...,n$ and $k=1- \mu \Delta t$.

\begin{remark}
The fixed points in eqn.(\ref{e42}) are identical to the fixed points of eqn.(\ref{e45}) or eqn.(\ref{e45a}). Moreover, the fixed point is independent of  $\Delta t$.
\label{rm41f}
\end{remark}

{\bf Proof of Remark \ref{rm41f}}.

Since the equations by setting $\frac{dy_{i}}{dt} =0$ in eqn.(\ref{e42}) are equivalent to the equations by $y_{i}(t+1) = y_{i}(t)$ in eqn.(\ref{e45}), they must have the same fixed points. 
   $\Box$

\begin{remark}
If $W=W^{T}$ and $ \mu >0$, eqn.(\ref{e45}) or eqn.(\ref{e45a}) coincides with a difference equational version of the Hopfield neural networks
\cite{Hopfield84,Hopfield85,Chen95b}.
\label{rm41}
\end{remark}

\subsection{Asymptotical stability of DRNN}

Similar to TCNN, we give several convergent theorems for  eqn.(\ref{e45a}), to show that the system of eqn.(\ref{e45a}) can asymptotically converge to a stable fixed point if certain conditions are satisfied.
Let $\lambda_{min}$ be the smallest eigenvalue of matrix $W$.

\begin{theorem}
\label{t41aa}
Assume $W=W^{T}$, and 
\begin{enumerate}
\item $1/3 \geq k \geq 0$,  $4(1-k) \epsilon > - \Delta t \lambda_{min}$ \ (or \ $4 \mu \epsilon > - \lambda_{min}$)\\
or 
\item $1 \geq k \geq 1/3$, $8k \epsilon > - \Delta t \lambda_{min}$\\
or 
\item $ k > 1$, $8 \epsilon > - \Delta t \lambda_{min}$.
\end{enumerate}
   Then  the discrete-time system of eqn.(\ref{e45a}) asymptotically converges to a fixed point, as far as eqn.(\ref{e45a}) is synchronously updated, where $k=1-\mu \Delta t$.
\end{theorem}

\begin{remark}
Assume $W=W^{T}$.
According to Gerschgorin Theorem (see Appendix \ref{appb}),  
\begin{enumerate}
\item  if $0 \leq k \leq 1/3$; $ \Delta t \omega_{ii} + 4(1-k) \epsilon  > \Delta t |\sum_{j \neq i}^{n} \omega_{ij}|$ for $i=1,...,n$, then the matrix $\Delta t W+ 4(1-k) \epsilon { \mbox{ \boldmath$1$}}$ is positive definite;\\

\item if $1/3 \leq k \leq 1$; $  \Delta t \omega_{ii} + 8k \epsilon > \Delta t |\sum_{j \neq i}^{n} \omega_{ij}|$ for $i=1,...,n$, then the matrix $\Delta t W+ 8k \epsilon { \mbox{ \boldmath$1$}}$ is positive definite;\\

\item if $1 < k $; $ \Delta t \omega_{ii} + 8 \epsilon > \Delta t |\sum_{j \neq i}^{n} \omega_{ij}|$ for $i=1,...,n$, then the matrix $\Delta t W+ 8 \epsilon { \mbox{ \boldmath$1$}}$ is positive definite.
\end{enumerate}
 These show that the system of $X(t)$ is asymptotically stable. 
\label{rm43a}
\end{remark}

\begin{theorem}
\label{t41a}
Assume $W=W^{T}$,
\begin{enumerate}
\item $1/3 \geq k \geq 0$,  $4(1-k) \epsilon > - \Delta t \ min_{i=1}^{n} \omega_{ii} $ \\
or 
\item $1 \geq k \geq 1/3$, $8k \epsilon > - \Delta t \ min_{i=1}^{n} \omega_{ii} $\\
or 
\item $ k > 1$, $8 \epsilon > - \Delta t \ min_{i=1}^{n} \omega_{ii} $.
\end{enumerate}
   Then the discrete-time system of eqn.(\ref{e45a}) asymptotically converges to a fixed point, as far as eqn.(\ref{e45a}) is asynchronously updated, where $k=1- \mu \Delta t$.
\end{theorem}

Define 
\begin{equation}
f(X)=-\frac{1}{2}\sum^{n}_{ij} \Delta t \omega_{ij}x_{i}x_{j} - \sum^{n}_{i} \Delta t a_{i} x_{i}
 - (k-1)\epsilon \sum^{n}_{i} \int_{0}^{x_{i}} ln \frac{x}{1-x} dx
\label{ely2}
\end{equation}

Then, Theorems \ref{t41aa} and \ref{t41a} can be proved as the same procedures of Theorems \ref{t31aa} and \ref{t31a}, by showing that $f(X)$ is a  Lyapunov function of eqn.(\ref{e45a}) for both synchronously and asynchronously updating.

\begin{remark}
 Theorem \ref{t41aa} and Theorem \ref{t41a} ensure that asymptotically stable points of discrete-time recurrent neural networks, i.e., eqn.(\ref{e45}) or eqn.(\ref{e45a}) are identical to those of continuous-time recurrent neural networks, i.e., eqn.(\ref{e42}) with $W=W^{T}$.
\end{remark}

\begin{remark}
 If $k= 0$ and $W=W^{T}$, the attractors of eqn.(\ref{e45a}) for both synchronously and asynchronously updating, include only fixed points and 2-periodic points \cite{Marcus} regardless of any values of $\omega_{ij}$ and $\Delta t$.    
\label{rm44a}
\end{remark}

This fact can be readily shown by use of the results of Marcus and Westervelt \cite{Marcus}.

\begin{lemma}
\label{l41aa}
Assume $W=W^{T}$.
\begin{enumerate}
\item Fixed points of $X(t)$ for eqn.(\ref{e45a}) are composed of stationary points of $f(X)$ (i.e., all of local minima, local maxima, saddle points of $f(X(t))$) and all of the vertices of $X(t)$, regardless of synchronous or asynchronous updating. 
\item As far as $k \neq 1$ or $\mu \neq 0$, there is no stationary points of $f(X)$ on the vertices or boundary points of $X$.
\end{enumerate}
\end{lemma}

\begin{theorem}
\label{t42a1a}
Assume the same conditions as Theorem \ref{t41aa} or Theorem \ref{t41a}.
Then a point is  asymptotically stable for eqn.(\ref{e45a}) if and only if this point is a local minimum of $f(X)$ for $E_{0} \leq X \leq E_{1}$.
\end{theorem}

\begin{corollary}
\label{r43a1}
Assume the same conditions as Theorem \ref{t41aa} or Theorem \ref{t41a} and assume  $\omega_{ii}=0$ for $i=1,...,n $ and $\mu =0$,    then a point  $X(t)$ is  asymptotically stable for eqn.(\ref{e45a}) if and only if this point is a local minimum of $f(X)$ for $E_{0} \leq X \leq E_{1}$ on the vertices.
\end{corollary}

Although Theorem \ref{t41aa} and Theorem \ref{t41a} are simple and easy to apply, they are  conservative estimators for the evaluation of asymptotical stability. Next, we give two more general theorems to evaluate the local stability in an exact manner in a similar way as Section 3.

For synchronous updating, define the Jacobian matrix $J_{X}$ of eqn.(\ref{e45a}) at $X(t)$ as;

\begin{eqnarray}
J_{X} = \frac{dX(t+1)}{dX(t)}  = k
 \left[
\begin{array}{ccc}
\frac{x_{1}(t+1)[1-x_{1}(t+1)]}{x_{1}(t)[1-x_{1}(t)]} &   & \bigzerou \\
 &   \ddots &                 \\
\bigzerou &         & \frac{x_{n}(t+1)[1-x_{n}(t+1)]}{x_{n}(t)[1-x_{n}(t)]}  \\
\end{array}
\right]    \label{e42a101} \\
 + \frac{1}{\epsilon}
 \left[
\begin{array}{ccc}
x_{1}(t+1)[1-x_{1}(t+1)] &   &  \bigzerou\\
 &   \ddots &                 \\
\bigzerou &          & x_{n}(t+1)[1-x_{n}(t+1)]  \\
\end{array}
\right] \Delta t W.
\label{e42a55}
\end{eqnarray}

Then for any $x_{i}=0$ or $1$,

\begin{eqnarray} 
\lim_{x_{i}(t) \rightarrow 0} \frac{x_{i}(t+1)}{x_{i}(t)} = 
 \left\{
\begin{array}{cc}
\infty  &  \mbox{if} \ k < 1 \\
e^{ \frac{1}{\epsilon}[\sum_{j=1}^{n} \Delta t \omega_{ij}x_{j}(t) + \Delta t a_{i}]}     &  \mbox{if} \ k = 1 \\
0       &  \mbox{if} \ k > 1 \\
\end{array}
\right.
\label{er41}
\end{eqnarray}

\begin{eqnarray} 
\lim_{x_{i}(t) \rightarrow 1} \frac{1-x_{i}(t+1)}{1-x_{i}(t)} = 
 \left\{
\begin{array}{cc}
\infty  &  \mbox{if} \ k < 1 \\
e^{- \frac{1}{\epsilon}[\sum_{j=1}^{n} \Delta t \omega_{ij}x_{j}(t) + \Delta t a_{i} ]}    &  \mbox{if} \ k = 1 \\
0       &  \mbox{if} \ k > 1 \\
\end{array}
\right.
\label{er42}
\end{eqnarray}

Therefore, the Jacobian matrix for a fixed point $X^{*}=(x_{1}^{*},...,x_{n}^{*})$ is 

\begin{equation}
J_{X}= diag(k_{1},...,k_{n})+ \frac{\Delta t}{\epsilon} diag[ x_{1}(1-x_{1}),..., x_{n}(1-x_{n})] W
\label{eqj11}
\end{equation}

where

\begin{eqnarray}
 k_{i} = 
 \left\{
\begin{array}{cc}
\infty  &   \mbox{if} \ k<1, x_{i}^{*}=0 \ \mbox{or} \ 1 \\
e^{ \frac{1-2x_{i}^{*}}{\epsilon}[\sum_{j=1}^{n} \Delta t \omega_{ij}x_{j}^{*} + \Delta t a_{i}]}     &  \mbox{if} \ k = 1, x_{i}^{*}=0 \ \mbox{or} \ 1 \\
0       &  \mbox{if} \ k > 1, x_{i}^{*}=0 \ \mbox{or} \ 1 \\
k       &   \mbox{if} \ x_{i}^{*} \neq 0 \ \mbox{or} \ \neq 1 \\
\end{array} 
\right\} . \label{eqj22}
\end{eqnarray}

Next, we give a theorem generalized to boundary points and vertices and internal points of $X$.

\begin{theorem}
\label{t42a11}
Assume $W=W^{T}$.

\begin{enumerate}
\item  For synchronous updating, a fixed point $X^{*}$ in the region $S$ is asymptotically stable if the largest eigenvalue $ \lambda_{max}^{'} $ and smallest eigenvalue $ \lambda_{min}^{'} $ of the matrix \ 
$J_{X}$ \ satisfy \ \ \ 
 $ \lambda_{max}^{'}  <1$ \ and \ $ \lambda_{min}^{'} >- 1$; \ \ \ 
on the other hand, a point in the region $S$ is unstable if either  \ 
 $ \lambda_{max}^{'}  >1$ \ or \ $ \lambda_{min}^{'} < -1$ \ is satisfied; 
where, $J_{X}$ is defined in eqn.(\ref{eqj11}).

\item  For asynchronous updating,  a fixed point in the region $S$ is asymptotically stable if \ 
$ 1 > \ max_{i=1}^{n} [k_{i} + \frac{\Delta t \omega_{ii} }{\epsilon } x_{i}^{*}(1-x_{i}^{*})] $ \ and \ $ min_{i=1}^{n} [k_{i} + \frac{\Delta \omega_{ii} }{\epsilon } x_{i}^{*}(1-x_{i}^{*})] > -1$.
\end{enumerate}
\end{theorem}

{\bf Proof of Theorem \ref{t42a11} }.

{\sl For synchronous updating.}

The Jacobian matrix is given in eqn.(\ref{eqj11}), and all eigenvalues of $J_{X}$ are real numbers according to Theorem \ref{appt1} of Appendix \ref{appc}.

{\sl For asynchronous updating}:
Without loss of generality, let neuron-$l$ be the updated neuron at iteration-(t+1), then 

 \begin{equation}
\frac{dx_{l}(t+1)}{dx_{l}(t)} = k
 \frac{x_{l}(t+1)[1-x_{l}(t+1)]}{x_{l}(t)(1-x_{l}(t))} 
 + 
\frac{ \Delta t \omega_{ll} }{\epsilon}
 x_{l}(t+1)[1-x_{l}(t+1)] 
\label{e42a12}
\end{equation}

Analogously to the proof of Theorem \ref{t32a11}, we obtain conditions $1$ and $2$.
  $\Box$

Let  $U_{a}$,  $D_{a}$ and  $L_{a}$  be an upper triangular, a diagonal and a lower triangular parts of $J_{X}$ defined in eqn.(\ref{e42a101}), i.e., $J_{X} = U_{a} + D_{a} + L_{a}$.
Define $\lambda_{cmin}, \lambda_{cmax}$ to be the minimal and maximal eigenvalues of matrix $({ \mbox{ \boldmath$1$}}-L_{a})^{-1}(U_{a} + D_{a})$, respectively. Then, let us show a stricter theorem for cyclic updating.

\begin{theorem}
\label{t42a1102}
Assume $W=W^{T}$. For cyclic updating,

\begin{enumerate}
\item  a fixed point $X^{*}$ in the region $S$ is asymptotically stable if the largest eigenvalue $ \lambda_{cmax} $ and smallest eigenvalue $ \lambda_{cmin}$ of the  matrix
 \ $({ \mbox{ \boldmath$1$}}- L_{a})^{-1} ( U_{a} + D_{a})$ \ satisfy \ \ \ 
 $| \lambda_{cmax}| <1$ \ and \ $| \lambda_{cmin} |< 1$;
\item  on the other hand, a fixed point $X^{*}$ in the region $S$ is unstable if either  \ 
 $| \lambda_{cmax} | >1$ \ or \ $| \lambda_{cmin} |> 1$ is satisfied.

\end{enumerate}
\end{theorem}

\begin{remark}
Assume $W=W^{T}$, then in the internal points of $X$, only local minima of $f(X)$ can possibly be asymptotically stable for eqn.(\ref{e45a}), regardless of synchronous or asynchronous updating.
\end{remark}

According to eqn.(\ref{eqj11}), the  closer $x_{i}^{*} \ (i=1,...,n)$ are  to $0$ and $1$, the smaller the absolute values of eigenvalues $\lambda_{min}^{'}$ and $\lambda_{max}^{'}$  are. Therefore, this Theorem implies that the system is likely to asymptotically converges to local minima near the vertices for $-1 <1- \mu \Delta t < 1$.

\begin{remark}
Assume $W=W^{T}$.

 \begin{enumerate}
\item When $k < 1$, no point on vertices or boundary points is asymptotically stable.  
 \item When $k > 1$, all vertices are asymptotically stable for synchronous and asynchronous updating. 
 
 \item For both synchronous updating and cyclic updating, when $k=1$, a vertex $X^{*}$ is asymptotically stable if  \ 
   $max_{i=1}^{n} e^{ \frac{1-2x_{i}^{*}}{\epsilon}[\sum_{j=1}^{n} \Delta t \omega_{ij}x_{j}^{*} + \Delta t a_{i} ]}   <1$; \ 
 on the other hand, a vertex is unstable if  \
    $max_{i=1}^{n} e^{ \frac{1-2x_{i}^{*}}{\epsilon}[\sum_{j=1}^{n} \Delta t \omega_{ij}x_{j}^{*} + \Delta t a_{i} ]}   >1$. \ 
 
 \item For random-order updating, when $k=1$, a vertex is asymptotically stable if  \

   $max_{i=1}^{n} e^{ \frac{1-2x_{i}^{*}}{\epsilon}[\sum_{j=1}^{n} \Delta t \omega_{ij}x_{j}^{*} + \Delta t a_{i} ]}   <1$. 

 \end{enumerate}
\end{remark}

Lemma \ref{l41aa}, Theorem \ref{t42a1a}, Remark \ref{r43a1} and Theorem \ref{t42a11}  can be proved as the same procedures of Lemma \ref{l31aa}, Theorem \ref{t32a1a}, Remark \ref{r33a1} and Theorem \ref{t32a11}.  Actually, by replacing  $\Delta t W$ and $ \Delta t I$ in eqn.(\ref{e45}) with  $W$ and $ I$ in eqn.(\ref{e5}) and further letting $I_{0} =E_{0}$ in eqn.(\ref{e5}), eqn.(\ref{e5}) coincides with eqn.(\ref{e45}). Therefore, all of theoretical results in this section can be proven as the same procedure of section $3$, only by taking these correspondences  into consideration.     

Theorems \ref{t42a1a}-\ref{t42a1102} are useful for both designs of neural networks and applications of DRNN to practical problems, e.g., optimization or associative memory, because they indicate which local minimum of Lyapunov function is a attractor and which one is not. 
In addition, the bifurcations of DRNN can also be analysed similarly as like Lemma \ref{lb1} and Theorems \ref{tb1},\ref{tb2} of TCNN. Actually, Lemma \ref{lb1}, Theorem \ref{tb1}, Theorem \ref{tb2}, Remark \ref{rb1} and Remark \ref{rb2} also hold for DRNN or eqn.(\ref{e45}) if we use $f$ defined in eqn.(\ref{ely2}) instead of $f$ defined in eqn.(\ref{ely}).

\begin{remark}
 Lemma \ref{lb1}, Theorem \ref{tb1}, Theorem \ref{tb2}, Remark \ref{rb1} and Remark \ref{rb2} also hold for DRNN or eqn.(\ref{e45}), where $f$ is defined in eqn.(\ref{ely2}).
\end{remark}

\clearpage
\pagestyle{myheadings}
\markright{6 CHAOS IN DRNN}

\section{Chaos in  Discrete-time Recurrent Neural Networks}

In this section, one theorem for a fixed point with sufficiently large difference times is firstly addressed, and then our main theorem for chaos is described in the last of this section. The function $f$ is defined in eqn.(\ref{ely2}). All the theorems in this section can be proved by similar arguments as in section 4.

\subsection{Fixed point in DRNN}
First, we show that eqn.(\ref{e45}) possesses a fixed point for sufficiently large $ \Delta t$.

\begin{theorem}
\label{t41}
Assume the following two conditions.
\begin{enumerate}
\item $W$ is invertible and $E_{1} > - W^{-1} I> E_{0}$.
\item $0 \leq \mu < d_{0}/ \Delta t$ , where $d_{0} \  (>0)$ is an arbitrary bounded number.
\end{enumerate}

Then there exists a positive constant $c_{13}$ such that for any $\Delta t > c_{13}$ the discrete-time system of eqn.(\ref{e45}) has one fixed point  $Y^{*}$ which is bounded. Furthermore $Y^{*}$ is a unique fixed point if $\Delta t $ is sufficiently large, and  $Y^{*}_{\infty} = \lim_{ \Delta t \rightarrow \infty} Y^{*}= \{ Y | X(Y) = -W^{-1}I  \} $.
\end{theorem}

{\bf Proof of Theorem \ref{t41}. }
This Theorem can be proved by a similar procedure  as in the proof of Theorem \ref{t31}.     $\Box$

Since $f_{XX} \propto - \Delta t W$ if $\Delta t$ is sufficiently large, we have the following remark. 

\begin{remark}
$Y^{*}$ or $X(Y^{*})$ is a unique maximum, saddle point or minimum of $f$ defined in eqn.(\ref{ely2}) if $W$ is (semi-) positive definite, indefinite or (semi-) negative definite, when $\Delta t$ is sufficiently large.  
\end{remark}

\subsection{Existence of chaos in DRNN}
Next, we establish a principal theorem of this section.

\begin{theorem}
\label{t42}
Assume the same conditions as Theorem \ref{t41}, and 
\begin{enumerate}
\item $k(E_{1}+W^{-1}I) > -W^{-1}I$; $I > E_{0}$,\\
or 
\item $ I <E_{0}, 0>k>-1$,\\
or 
\item $kW^{-1}I <-(W^{-1}I+ E_{1})$, $W E_{1} + I <E_{0}$,\\
or 
\item $W E_{1} + I >E_{0}, 0>k>-1$.
\end{enumerate}

then there exists a positive constant $c_{14}$ such that for any $\Delta t> c_{14}$ the discrete-time system of eqn.(\ref{e45}) is chaotic in the sense of Marotto.
\end{theorem}

\begin{lemma}[Hata, 1982]
\label{l41}
There exists a positive constant $c_{15}$ such all eigenvalues of $D_{Y}f(Y)$ exceed the unity in norm for all $\Delta t > c_{15}$ and for any bounded $Y$.
\end{lemma}

{\bf Proof of Lemma \ref{l41}}. 
This lemma can be proved by a similar argument as in the proof of Hata's Lemma \cite{Hata}.     $\Box$

{\bf Proof of Theorem \ref{t42}}. 
Define $[H_{1},...,H_{n} ]$ as
\begin{equation}
[H_{1},...,H_{n}]^{T} = W^{-1}
  \label{e65}
\end{equation}

{\sl For conditions {\it 1 } and {\it 2} of Theorem \ref{t42}}:

 Let $G=(g_{1},...,g_{n})^{T}$, where, 

\begin{equation}
g_{i} = \frac{H_{i}^{T}}{\Delta t} 
[ 
\frac{Y^{*} - \Delta t I}{k} - \Delta t I - \epsilon k
\left[
\begin{array}{c}
ln \frac{-H_{1}^{T}I(1+1/k)}{1+H_{1}^{T}I(1+1/k)}   \\
\vdots   \\
ln \frac{-H_{n}^{T}I(1+1/k)}{1+H_{n}^{T}I(1+1/k)} 
\end{array}
\right]  ]
\label{e47}
\end{equation}

Then define
\begin{equation}
 Y^{0}  = \epsilon [ln \frac{g_{1}}{1-g_{1}},..., ln \frac{g_{n}}{1-g_{n}}]^{T}. \label{e48}
\end{equation} 

Therefore, $X(Y^{0})= G$ by eqns.(\ref{e45}),(\ref{e47}) and (\ref{e48}). The following condition should be satisfied  in eqn.(\ref{e48}).

\begin{equation}
0 < g_{i} < 1  \ \ (i=1,..,n)                            \label{e50} 
\end{equation}

Note that $ \lim_{\Delta t \rightarrow \infty} g_{i} = - H_{i}^{T}I(1/k+1)$ which is between $0$ and $1$ ensured by the condition $1$ of Theorem \ref{t41} and the conditions $1$ and $2$ of Theorem \ref{t42}.

Since,
\begin{equation}
F(Y^{0}) = \frac{1}{k}(Y^{*}- \Delta t I) +k \epsilon  
\left[
\begin{array}{c}
ln \frac{g_{1}}{1-g_{1}} - ln \frac{-H_{1}^{T}I(1+1/k)}{1+H_{1}^{T}I(1+1/k)} \\
\vdots   \\
ln \frac{g_{n}}{1-g_{n}} - ln \frac{-H_{n}^{T}I(1+1/k)}{1+H_{n}^{T}I(1+1/k)}
\end{array}
\right],
\label{e51}
\end{equation}

then for sufficiently large $\Delta t$
\begin{equation}
F(Y^{0}) \propto -\frac{\Delta t I}{k}
\label{e270a}
\end{equation}
 which is negative because of the conditions $1$ and $2$.

{\sl For conditions {\it 3} and {\it 4} of Theorem \ref{t42}}:

 Let $G=(g_{1},...,g_{n})^{T}$, where, 

\begin{equation}
g_{i} = \frac{H_{i}^{T}}{\Delta t} 
[ 
\frac{Y^{*} - \Delta t(I+WE_{1})}{k} -\Delta t I - \epsilon k
\left[
\begin{array}{c}
ln \frac{-H_{1}^{T}I(1+1/k) -1/k}{1+ H_{1}^{T}I(1+1/k)+1/k}   \\
\vdots   \\
ln \frac{-H_{n}^{T}I(1+1/k) -1/k}{1+ H_{n}^{T}I(1+1/k)+1/k} 
\end{array}
\right]   ]
\label{e77}
\end{equation}

Then define
\begin{equation}
 Y^{0} = \epsilon [ln \frac{g_{1}}{1-g_{1}},...,ln \frac{g_{n}}{1-g_{n}}]^{T}.
\label{e78}
\end{equation} 

Therefore, $X(Y^{0})= G$. The following  condition should be satisfied  in eqn.(\ref{e78}).

\begin{equation}
0 < g_{i} < 1   \ \ (i=1,..,n)                            \label{e501} 
\end{equation}

Note that $\lim_{\Delta t \rightarrow \infty} g_{i} =- H_{i}^{T}I(1/k+1)-1/k$ which is between $0$ and $1$  ensured by the condition $1$ of Theorem \ref{t41} and the conditions $3$ and $4$ of Theorem \ref{t42}.

Since
\begin{equation}
F(Y^{0}) = \frac{1}{k}[Y^{*}- \Delta t( I+WE_{1})] +k \epsilon  
\left[
\begin{array}{c}
ln \frac{g_{1}}{1-g_{1}} - ln \frac{-H_{1}^{T}I(1+1/k)-1/k}{1+H_{1}^{T}I(1+1/k)+1/k}   \\
\vdots   \\
ln \frac{g_{n}}{1-g_{n}} - ln \frac{-H_{n}^{T}I(1+1/k)-1/k}{1+H_{n}^{T}I(1+1/k)+1/k} 
\end{array}
\right],
\label{e51a}
\end{equation}

then for sufficiently large $\Delta t$
\begin{equation}
F(Y^{0}) \propto -\frac{\Delta t (I+WE_{1})}{k}
\label{e272}
\end{equation}
 which is positive because of the conditions $3$ and $4$ of Theorem \ref{t42}.

Then Theorem \ref{t42} can be proved as the same procedure as that of proof of Theorem \ref{t32}.     $\Box$ 

Obviously, there are two different type of snap-back repellers for Theorem \ref{t42}.   One  is $ \{ Y^{0*} \Longrightarrow  \ \mbox{(sufficiently negative)} \ F(Y^{0*})  \Longrightarrow  Y^{*} \} $ corresponding to conditions $1$ and $2$, the other is $ \{ Y^{0*} \Longrightarrow  \ \mbox{(sufficiently positive)} \ F(Y^{0*}) \Longrightarrow  Y^{*} \} $ corresponding to conditions $3$ and $4$.

\begin{remark}
Chaos in eqn.(\ref{e45}) is generated not from a saddle point, but from a repeller when $\Delta t$ is sufficiently large.
\label{rm43}
\end{remark}

\begin{remark}
When $\Delta t$ is sufficiently large, eqn.(\ref{e45}) can generate chaotic behaviour even if the system is symmetry without self-feedback weights, i.e., $W=W^{T}$ and $\omega_{ii}=0; i=1,...,n$.
\label{rm44}
\end{remark}

\begin{theorem}
 $Y(t)$ of eqn.(\ref{e45}) is bounded for any initial value $Y(0)$ if $-1 < k < 1$.
\label{rm45}
\end{theorem}

\begin{example}
\label{ex41}
 Consider a one-dimensional case of eqn.(\ref{e45}),

\[ y(t+1) = (1- \mu \Delta t)y(t)+ \Delta t ( \omega x(y(t)) +a) \ ; \ x(y(t))=\frac{1}{1+e^{-y(t)/ \epsilon }}. \]

Let $\mu =1; \omega =-1; \epsilon =1/250; a=0.4$. Figure \ref{f2} shows the output of neuron and the maximum Lyapunov exponent with increasing $\Delta t$. 
\end{example}

Analogously to Example \ref{ex31}, we first use the condition $2$ of Theorem \ref{t41aa} or Theorem \ref{t41a} to estimate the stability of this example.
 According to the condition $2$ of Theorem \ref{t41a}, the estimated value of $\Delta t $ is $\Delta t_{c}= \frac{8 \epsilon }{8 \epsilon \mu -\omega} =  0.03101$, below which the system can be ensured to be stable. If $\Delta t  < 0.03101$, the system is stable and converges to a fixed point, which is verified by the numerical simulation in Figure \ref{f2}. Since the Theorem \ref{t41aa} and Theorem \ref{t41a} are conservative although they are easy to use, next we apply Theorem \ref{t42a11} to exactly evaluate locally asymptotical stability of this example.

 The fixed point satisfies

\begin{equation}
- \mu \epsilon ln \frac{x(t)}{1-x(t)}+ \omega x(t) +  a =0,
\label{ee21}
\end{equation}
independent of $\Delta t$.

According Theorem \ref{t42a11}, the  condition under which the fixed point is stable is as follows

\begin{equation}
k + \frac{\Delta t}{\epsilon} \omega x(t)(1-x(t)) > -1, 
\label{ee22}
\end{equation}

because we only analyse the performances for $k>0$.
Therefore, by solving eqns.(\ref{ee21}) and (\ref{ee22}), we obtain the critical values of $\Delta t$ and fixed point $x^{*}$, 

\[\Delta t_{c} = 0.032744 ; x^{*c} = 0.401595. \]

Therefore,  If $ \Delta t < 0.032744$, the system is stable and converges to a fixed point; at $\Delta t =0.032744$, the fixed point bifurcates into 2-periodic points;  when $ \Delta t > 0.032744$, the system becomes unstable.  These facts are verified by the numerical simulation in Figure \ref{f2}.

Finally, we use the  condition $1$ of Theorem \ref{t42} to analyze chaotic dynamics of this example. 
 According to the condition $1$ of Theorem \ref{t42}, there exists chaotic behavior when $\Delta t $ is sufficiently large.  Actually, in this example, when $ \Delta t > 0.09536$, the chaotic behavior appears, which is numerically verified by positive maximum Lyapunov exponents in Figure \ref{f2}, although there also exist many periodic windows.

\begin{example}
\label{ex42}
 Consider a two-dimensional case of eqn.(\ref{e45}).
Let $\mu =1; \omega_{11}= -1; \omega_{22} =-2; \omega_{12}=-0.5;
\omega_{21}= -0.5; \epsilon =1/250; a_{1}=a_{2}=0.2$. Figures \ref{f3}
and \ref{f4} show the outputs of neurons and the maximum Lyapunov exponent with increasing $\Delta t$.  
\end{example}

Since the minimal and maximal eigenvalues of $W$ are $\lambda_{min} = \frac{-3- \sqrt{2}}{2}$ and $\lambda_{max} = \frac{-3+ \sqrt{2}}{2}$, the condition $2$ of Theorem \ref{t41aa} or Theorem \ref{t41a} can be used to analyse asymptotical stability for $\Delta t < \frac{2}{3 \mu }=0.67$.

{\sl For the synchronous updating}: According to the condition $2$ of Theorem \ref{t41aa}, the estimated  value of $\Delta t $ is $\Delta t_{e}= \frac{8 \epsilon }{8 \epsilon \mu -\lambda_{min}} =  0.01429$, below which the system can be ensured stable.  Actually, as shown in Figure \ref{f3}, when $ \Delta t < 0.0416$, the system is stable and converges to a fixed point. Next, we use Theorem \ref{t42a11} to analyses local stability of this example in more exact manner. 
 The fixed point satisfies

\begin{equation}
- \mu \epsilon ln \frac{x_{i}(t)}{1-x_{i}(t)}+ \sum_{j=1}^{2} \omega_{ij} x_{j}(t) +  a_{i} =0, \ (i=1,2)
\label{ee31}
\end{equation}
independent of $\Delta t$.
The fixed point is $X^{*}=(x_{1}^{*},x_{2}^{*})= (0.175392,0.0615991)$ by solving eqns.(\ref{ee31}). 
According to the condition $1$ of Theorem \ref{t42a11}, the condition under which the fixed point is stable follows

\begin{equation}
1- \mu \Delta t + \frac{\Delta t \lambda_{min}^{''}}{\epsilon} >-1, \ \mbox{and} \  1- \mu \Delta t + \frac{\Delta t \lambda_{max}^{''}}{\epsilon}  <1,
\label{ee32}
\end{equation}
where eigenvalues of $diag[ x_{1}^{*}(1-x_{1}^{*}),..., x_{n}^{*}(1-x_{n}^{*})] W$ are calculated to be $\lambda_{min}^{''}= -0.17827; \lambda_{max}^{''} = -0.08230$.
Therefore, by solving eqns.(\ref{ee32}), we obtain the critical values of $\Delta t$, 

\[\Delta t_{c} = 0.04381 . \]

 If $ \Delta t < 0.04381$, the system is stable and converges to a fixed point; at $\Delta t =0.04381$, the fixed point bifurcates into 2-periodic points;  when $ \Delta t > 0.04381$, the system becomes unstable.  These facts are verified by the numerical simulation in Figure \ref{f3}.

{\sl For the asynchronous updating}: Here, we use cyclic updating for simulation. According to the condition $2$ of Theorem \ref{t41a}, the estimated value of $\Delta t $ is $\Delta t_{e}= \frac{8 \epsilon }{8 \epsilon \mu -\omega_{ii}} = 0.01575$, below which the system can be ensured to be stable.  Actually, as shown in Figure \ref{f4}, when $ \Delta t < 0.01575$, the system is stable and converges to a fixed point. 
 Next, we use Theorem \ref{t42a11} to analyse stability of this example in more exact manner. 
 Since the fixed point is independent of $\Delta t$ and updating schemes, 
the fixed point is $X^{*}=(x_{1}^{*},x_{2}^{*})= (0.175392,0.0615991)$  calculated in eqns.(\ref{ee31}). 
According to the condition $2$ of Theorem \ref{t42a11}, the sufficient condition under which the fixed point is locally stable follows

\begin{equation}
1- \mu \Delta t + \frac{\Delta t \omega_{ii}}{\epsilon} x_{i}^{*}(1-x_{i}^{*})  > -1, \ \mbox{for} \ i=1,2,
\label{ee42}
\end{equation}
because of $\mu >0$ and $\omega_{ii}<0$. 
Therefore, by solving eqns.(\ref{ee42}), we obtain, 

\[\Delta t_{s} = 0.053825 . \]

 If $ \Delta t < 0.053825$, the system is stable and converges to a fixed point.  When $ \Delta t > 0.0582$, the system becomes unstable, as shown by the numerical simulation in Figure \ref{f4}. 

On the other hand,  according to the condition $1$ of Theorem \ref{t42}, there exists chaotic behavior when $ \Delta t $ is sufficiently large.  Actually, in this example, when $ \Delta t > 0.06066$ (for synchronous updating) or $\Delta t > 0.06915$ (for asynchronous updating), the chaotic behavior appears, which is numerically verified by positive maximum Lyapunov exponents in Figures \ref{f3} and \ref{f4}.

\clearpage
\pagestyle{myheadings}
\markright{7 CONCLUSION}

\section{Conclusion}

This paper theoretically proved that both TCNN and difference equations of discrete-time recurrent neural networks (DRNN) have chaotic structure by applying Marotto' Theorem, and gave sufficient conditions for existence of both a fixed point and chaotic behavior. A significant characteristic of TCNN and DRNN  is that they have only one fixed point which generates chaotic behaviour of neural networks when the absolute value of self-feedback connection weights in TCNN and the difference time in DRNN are sufficiently large. Besides, this paper also 
investigated the relations between the asymptotically stable points and the minimal solutions of the Lyapunov function, and 
derived the  conditions for which both TCNN and DRNN can asymptotically converge to a fixed point,  which ensure the convergence of {\sl chaotic simulated annealing}\cite{Chen95a,Chen95b}. These convergent theorems are also useful for applications of TCNN and DRNN to practical problems, e.g., optimization and associative memory, because they indicate which local minimum of Lyapunov function is an attractor and which one is not.  The local bifurcations were also investigated to show  the bifurcation process for chaotic simulated annealing \cite{Chen95b}. Since the models which we used in this paper are simple and general, the obtained theoretical results hold for a wide class of discrete-time neural networks.

\clearpage
\pagestyle{myheadings}
\markright{REFERENCE}

\clearpage
\pagestyle{myheadings}
\markright{APPENDIX}

\appendix

\begin{center}\LARGE
{\bf APPENDICES}
\end{center}

\vspace*{2cm}

\section{Urabe's Proposition (1965)}
\label{appa}

Let $F: \mbox{ \boldmath$  R $}^{n} \rightarrow \mbox{ \boldmath$  R $}^{n}$ be  a continuously differentiable function in some region $\Omega \subset \mbox{ \boldmath$  R $}^{n}$. Assume that equation 

\[ F(Y) = 0 \]

has an approximate solution $Y= \bar{Y}$  for which $det(F_{Y}(\bar{Y}))$ is not zero, and there is a constant $\epsilon \   (>0)$ and a constant $\mu \ (\geq 0)$ such that

\begin{enumerate}
\item $ \Omega_{\epsilon} =\{ Y | \| Y- \bar{Y} \| < \epsilon \} \subset \Omega $,
\item $\| F_{Y}(Y) -F_{Y}(\bar{Y}) \| < \mu /d$ for any $Y \subset \Omega_{\epsilon}$, \item $\frac{r d}{1- \mu} < \epsilon$,
\end{enumerate}
 where $r$ and $d \   (>0)$  are positive numbers such that $\| F(\bar{Y}) \| < r $ and $\| F_{Y}(\bar{Y})^{-1}\| < d$.

Then $F(X)=0$ has one and only one solution $Y=Y^{*}$ in $\Omega_{ \epsilon}$ and   $\| \bar{Y} - Y^{*} \| < \frac{r d}{1- \mu} $.\\[1.5cm]

\section{Gerschgorin's Theorem}
\label{appb}

Consider a matrix $W$

\begin{equation}
W =
  \left[
\begin{array}{ccc}
\omega_{11} & \cdots & \omega_{1n} \\ 
\vdots      & \vdots & \vdots      \\
\omega_{n1} & \cdots & \omega_{nn} 
\end{array}
\right] \nonumber
\end{equation}  

Let $\lambda_{p} \in \mbox{ \boldmath$  C $}; (p=1,...,n)$ be eigenvalues of $W$, and $\theta_{i} = \omega_{ii};
r_{i} = \sum_{j \neq i}^{n} | \omega_{ij}| ; (i=1,...,n)$.

Then $\lambda_{p} \in \bigcup_{i=1}^{n} B(\theta_{i};r_{i}) ; (p=1,...,n).$ In other words, 
there exists a set $S_{0} \subset \{1,...,n \}$ such that the following inequalities hold  for some $i \in S_{0}$.

\begin{equation}
r_{i} \geq | \theta_{i} - \lambda_{p} | \ \ \ \mbox{for} \ p=1,...,n \nonumber
\end{equation}

\vspace*{1.5cm}

\section{Some Facts for Matrices}
\label{appc}

A matrix $W$ and  two diagonal matrices $K$, $Z$ are defined as,

\begin{equation}
W =
  \left[
\begin{array}{ccc}
\omega_{11} & \cdots & \omega_{1n} \\ 
\vdots      & \vdots & \vdots      \\
\omega_{n1} & \cdots & \omega_{nn} 
\end{array}
\right] \ ; \  
K = diag(k_{11},...,k_{nn}) \ ; \  
Z = diag(z_{11},...,z_{nn}),
 \nonumber
\end{equation}  

where  $z_{ii} \geq 0$ \ for \ $i=1,...,n$. 
 Define a diagonal matrix $\sqrt{Z} = diag( \sqrt{z_{11}},..., \sqrt{z_{nn}})$.

\begin{lemma}
\label{appl1}
 The matrix $K + ZW$ has the same eigenvalues as those of the matrix $K + \sqrt{Z} W \sqrt{Z}$. 
\end{lemma}

{\bf Proof of Lemma \ref{appl1}}. 
 
Since $z_{ii} \geq 0$ \ for \ $i=1,...,n$, it is easy to show that algebraic equation \ $det(K + ZW - \lambda { \mbox{ \boldmath$1$}}) = 0$ for $\lambda $ \ is identical to \ $det(K + \sqrt{Z} W \sqrt{Z} - \lambda { \mbox{ \boldmath$1$}}) = 0$, which means that $K + ZW$ and $K + \sqrt{Z} W \sqrt{Z}$ have the same eigenvalues.  
  $\Box$

Then following results hold. 

\begin{theorem}
\label{appt1}
Assume $W$ is a symmetrical matrix, i.e., $W=W^{T}$, and $k_{ii},z_{ii}$ are real numbers. Then the following results hold.

\begin{enumerate}
\item The matrix $K + ZW$ has the same eigenvalues as those of the matrix $K + \sqrt{Z} W \sqrt{Z}$. All of these eigenvalues are real numbers.

\item Let $z_{ii} >0$ for $i=1,...,n$.  Then properties of matrix $\sqrt{Z} W \sqrt{Z}$ are identical to those of $W$ in terms of positive definite (or semi-positive definite), indefinite and negative definite (or semi-negative definite).  All eigenvalues for $\sqrt{Z} W \sqrt{Z}$ (or $ZW$) are real numbers.

\item Let $z_{ii} > 0$ for $i=1,...,n$. Then all eigenvalues of $ZW$ is positive (or semi-positive), if and only if $W$ is postive definite (or semi-positive definite).  All eigenvalues of $ZW$ is negative (or semi-negative), if and only if $W$ is  negative definite (or semi-negative definite). Furthermore, $ZW$ has both positive and negative eigenvalues, if and only if $W$ is indefinite.
\end{enumerate}
\end{theorem}

{\bf Proof of Theorem \ref{appt1}.}

{\sl For condition $1$.}
 
According to Lemma \ref{appl1}, $K + ZW$ and $K + \sqrt{Z} W \sqrt{Z}$ have the same eigenvalues. 
 Since $K + \sqrt{Z} W \sqrt{Z}$ is a symmetrical matrix due to $W^{T} = W$, all eigenvalues of $K + \sqrt{Z} W \sqrt{Z}$ as well as $K + ZW$ are real numbers.

{\sl For condition $2$.}

Define a principal submatrix $W_{i}$ of $W$ as,

\begin{equation}
W_{i} =
  \left[
\begin{array}{ccc}
\omega_{11} & \cdots & \omega_{1i} \\ 
\vdots      & \vdots & \vdots      \\
\omega_{i1} & \cdots & \omega_{ii} 
\end{array}
\right] \ \mbox{for} \ i=1,...,n.
\nonumber
\end{equation}  

Let $A=\sqrt{Z}W \sqrt{Z}$, then 

\[ A_{i}= \sqrt{Z}_{i}W_{i} \sqrt{Z}_{i} \ \ \ (i=1,...,n), \]
where $\sqrt{Z}_{i}$ and $A_{i}$ are submatrices of $\sqrt{Z}$ and $A$ defined as the same as  $W_{i}$, respectively. Note that different from an asymmetrical matrix  $ZW$, $\sqrt{Z} W \sqrt{Z}$ or $A$ is a symmetrical matrix.

Let us first show that $W$ is positive definite if $A = \sqrt{Z}W \sqrt{Z} >0$.
The necessary and sufficient conditions for which $A$ is positive definite are
 
\[ det(A_{i})=det(\sqrt{Z}_{i}) det(W_{i}) det(\sqrt{Z}_{i}) >0  \ \ \ (i=1,...,n). \]

Since $z_{ii} >0$ for $i=1,...,n$ which means that $\sqrt{Z}$ is positive definite,  $det(\sqrt{Z}_{i}) >0$ for $i=1,...,n$. Therefore, to ensure $det(\sqrt{Z}_{i}) det(W_{i}) det(\sqrt{Z}_{i})$ $>0 $ for $ i=1,...,n$, the inequalities $det(W_{i}) >0 $ for $i=1,...,n$ must hold, which means that $W$ must be positive definite.

On the other hand, if $W > 0$,  then $det(\sqrt{Z}_{i}) det(W_{i}) det(\sqrt{Z}_{i})$ $>0$ for $i=1,...,n$, which means that $A= \sqrt{Z}W \sqrt{Z}$ is also positive definite. Therefore, the positive definiteness of $\sqrt{Z} W \sqrt{Z}$ is identical to that of $W$.

As the same way, we can prove the cases for semi-positive definite, indefinite,  negative definite and semi-negative definite.

{\sl For condition $3$.} 
Let $k_{ii}=0$ for $i=1,...,n$ in condition $1$, then from the conditions $1$ and $2$, the proof of condition $3$ is straightforward. 
    $\Box$

\clearpage
\pagestyle{myheadings}
\markright{Captions of Figure and Table}

\begin{center}
Captions of Figure and Table
\end{center}

\vspace*{1cm}

\begin{itemize}
\item Figure \ref{f1} : Output of the neuron and the Lyapunov exponent in the single neuron dynamics of TCNN with increasing $| \omega |$. $k=0.9; \epsilon =1/250; a=0; a_{0}= 0.65.$
 (a) Output of neuron and  (b) Lyapunov exponent.
\item Figure \ref{f2} : Output of the neuron and the Lyapunov exponent in the single neuron dynamics of DRNN with increasing $ \Delta t$. $\mu =1; \omega =-1; \epsilon =1/250; a=0.4.$
 (a) Output of neuron and (b) Lyapunov exponent.
\item Figure \ref{f3} : Outputs of the neurons and the maximum Lyapunov exponent in two-neuron dynamics of DRNN with synchronous updating. $\mu=1; \omega_{11}=-1; \omega_{22}=-2; \omega_{12}=-0.5; \omega_{21}=-0.5; \epsilon =1/250; a_{1}=a_{2}=0.2.$
 (a) Outputs of neurons and  (b) Maximum Lyapunov exponent.
\item Figure \ref{f4} : Outputs of the neurons and the maximum Lyapunov exponent in two-neuron dynamics of DRNN with asynchronous updating. $\mu=1; \omega_{11}=-1; \omega_{22}=-2; \omega_{12}=-0.5; \omega_{21}=-0.5; \epsilon =1/250; a_{1}=a_{2}=0.2.$
 (a) Outputs of neurons and  (b) Maximum Lyapunov exponent.
\end{itemize}

\clearpage
\pagestyle{myheadings}
\markright{Figure and Table}

\begin{figure}
\vspace*{12.5cm}
\caption{Output of neuron and the Lyapunov exponent in the single neuron dynamics of TCNN with increasing $| \omega |$}
\label{f1}
\end{figure}
   
\begin{figure}
\vspace*{12.5cm}
\caption{Output of neuron and the Lyapunov exponent in the single neuron dynamics of DRNN with increasing  $\Delta t$}
\label{f2}
\end{figure}
   
\begin{figure}
\vspace*{20.5cm}
\caption{Outputs of neurons and the maximum Lyapunov exponent in two-neuron dynamics of DRNN with synchronous updating}
\label{f3}
\end{figure}
   
\begin{figure}
\vspace*{20.5cm}
\caption{Outputs of neurons and the maximum Lyapunov exponent in two-neuron dynamics of DRNN with asynchronous updating}
\label{f4}
\end{figure}

\end{document}